\documentclass[sigconf]{acmart}
\AtBeginDocument{%
  }

\copyrightyear{2026}
\acmYear{2026}
\setcopyright{cc}
\setcctype{by}
\acmConference[CHI '26]{Proceedings of the 2026 CHI Conference on Human Factors in Computing Systems}{April 13--17, 2026}{Barcelona, Spain}
\acmBooktitle{Proceedings of the 2026 CHI Conference on Human Factors in Computing Systems (CHI '26), April 13--17, 2026, Barcelona, Spain}
\acmPrice{}
\acmDOI{10.1145/3772318.3790332}
\acmISBN{979-8-4007-2278-3/2026/04}

\usepackage{hyperref}
\usepackage{xspace}

\usepackage{booktabs}
\usepackage{todonotes}
\usepackage{enumitem}
\usepackage{multirow}
\usepackage{graphicx}
\usepackage{xspace}
\usepackage{fontawesome}
\usepackage{csquotes}

\newcommand{\eg}{{\it e.g.}, }
\newcommand{\ie}{{\it i.e.}, }

\newcommand{\vs}{{\it vs.}\xspace}

\newcommand{\one}[0] {{\it (i)}~}
\newcommand{\two}[0] {{\it (ii)}~}
\newcommand{\three}[0] {{\it (iii)}~}

\renewcommand{\arraystretch}{1.5}

\usepackage{multirow}
\usepackage{makecell}
\usepackage{color-edits}
\usepackage{graphicx}
\usepackage{subcaption}
\usepackage{tablefootnote}
\usepackage{longtable}
\PassOptionsToPackage{table,xcdraw}{xcolor}
\usepackage{colortbl}
\usepackage{ragged2e}
\usepackage{booktabs}  
\usepackage{array}     
\usepackage{ulem}

\definecolor{lightgray}{rgb}{0.83, 0.83, 0.83}
\definecolor{darkgreen}{rgb}{0.30, 0.82, 0.14}

\newcommand{\actionbox}{\colorbox{gray!20}{Action Sequence}}
\newcommand{\behaviorbox}{\colorbox{gray!20}{Behavior}}

\newcommand{\taskDurationWithUnit}{15 minutes}
\newcommand{\expOneCaps}{Quiz Solving}
\newcommand{\expTwoCaps}{Article Summarization}
\newcommand{\expThreeCaps}{Trip Planning}
\newcommand{\expOne}{\textit{quiz solving}}
\newcommand{\expTwo}{\textit{article summarization}}
\newcommand{\expThree}{\textit{trip planning}}
\newcommand{\withoutAI}{\textit{without LLM}}
\newcommand{\withAI}{\textit{with LLM}}

\newcommand{\participantNum}{77}
\newcommand{\hol}{users with high overreliance}
\newcommand{\holup}{Users with high overreliance}
\newcommand{\holSingle}{user with high overreliance}
\newcommand{\lol}{users with low overreliance}
\newcommand{\lolup}{Users with low overreliance}
\newcommand{\lolSingle}{user with low overreliance}
\newcommand{\llmres}{LLM’s responses}
\newcommand{\llmsres}{LLMs’ responses}
\newcommand{\llmresSingle}{LLM’s response}
\newcommand{\olm}{overreliance on LLMs}

\begin{document}

\title{Behavioral Indicators of Overreliance During Interaction with Conversational Language Models}

\author{Chang Liu}
\authornote{Both authors contributed equally to this research.}
\email{c-liu21@tsinghua.org.cn}
\orcid{0000-0002-1444-0993}
\affiliation{%
  \institution{Tsinghua University}
  \city{Beijing}
  \country{China}
}

\author{Qinyi Zhou}
\authornotemark[1]
\email{qinyi.zhou@connect.ust.hk}
\orcid{0009-0000-1878-3646}
\affiliation{%
  \institution{Hong Kong University of Science and Technology}
  \city{Hongkong}
  \country{China}
}

\author{Xinjie Shen}
\email{xinjie@gatech.edu}
\orcid{0009-0004-9176-5400}
\affiliation{%
  \institution{Georgia Institute of Technology}
  \city{Atlanta}
  \state{Georgia}
  \country{USA}
}

\author{Xingyu Bruce Liu}
\email{xingyuliu@ucla.edu}
\orcid{0000-0002-6988-5471}
\affiliation{%
  \institution{UCLA}
  \city{Los Angeles}
  \state{California}
  \country{USA}
}

\author{Tongshuang Wu}
\email{sherryw@cs.cmu.edu}
\orcid{0000-0003-1630-0588}
\affiliation{%
  \institution{Carnegie Mellon University}
  \city{Pittsburgh}
  \state{Pennsylvania}
  \country{USA}
}

\author{Xiang `Anthony' Chen}
\email{xac@ucla.edu}
\orcid{0000-0002-8527-1744}
\affiliation{%
  \institution{UCLA}
  \city{Los Angeles}
  \state{California}
  \country{USA}
}

\renewcommand{\shortauthors}{Liu, Zhou et al.}
\newcommand{\ai}{conversational LLMs\xspace}
\newcommand{\aiUp}{Conversational LLMs\xspace}

\begin{abstract}
  LLMs are now embedded in a wide range of everyday scenarios.
However, their inherent hallucinations risk hiding misinformation in fluent responses, raising concerns about overreliance on AI.
Detecting overreliance is challenging, as it often arises in complex, dynamic contexts and cannot be easily captured by post-hoc task outcomes.
In this work, we aim to investigate how users’ behavioral patterns correlate with overreliance.
We collected interaction logs from 77 participants working with an LLM injected plausible misinformation across three real-world tasks and we assessed overreliance by whether participants detected and corrected these errors. By semantically encoding and clustering segments of user interactions, we identified five behavioral patterns linked to overreliance: \lol{} show careful task comprehension and fine-grained navigation; \hol{} show frequent copy-paste, skipping initial comprehension, repeated LLM references, coarse locating, and accepting misinformation despite hesitation. We discuss design implications for mitigation.

\end{abstract}



\maketitle

\section{Introduction}
\aiUp{}, \eg ChatGPT \cite{gpt4}, Gemini \cite{gemini, gemini2}, and Claude \cite{claude}, are increasingly used for various open-ended tasks, from creative writing to problem-solving. These models hold promise for enhancing human productivity and creativity by providing knowledge-driven assistance. As a result, many users have come to rely on LLMs for reasons such as fast information access, reduced cognitive load, perceived authority, and lower social costs of seeking assistance. However, \llmsres{}, while often appearing fluent and plausible, may contain misinformation stemming from hallucinations, outdated information, and prompt injection attacks \cite{DBLP:journals/corr/abs-2404-18930, DBLP:journals/corr/abs-2401-11817, huang2025survey}. When users over-rely on LLMs and fail to critically evaluate their responses, such misinformation can lead to serious errors. Ultimately, this undermines the effectiveness of human-AI collaboration, reduces efficiency, and results in low-quality content~\cite{passi2024appropriate, weidinger_ethical_2021, thepriciple2019green}.

In this paper, we focus specifically on \olm{}, which is defined as \textbf{``users accepting incorrect LLM recommendations''}~\cite{passi2022overreliance, buccinca2021trustcognitiveforcing, jeon2025empowering, vasconcelos2023explanations, hunter2024monitoring}. Different from trust, which is an internal attitude that may lead to reliance \cite{howtoevaluate2021vereschak,hauptman2022componentstrust}
, reliance itself is the action taken by the human that can be observed and measured.
The conventional approach of understanding users' overreliance is through measuring and comparing the outcomes of tasks with \vs without AI assistance~\cite{bansal2021does,Gajos2021ToTOA,Inkpen2022AdvancingHCA}.

However, in the case of \ai, focusing solely on task outcomes overlooks the interaction process, which often spans multiple rounds of question-answering. While each round can be considered a task outcome in itself, evaluating individual rounds in open-ended tasks is difficult, as the content of each round may not directly relate to the final result. These intermediate rounds influence the final outcome, and errors caused by overreliance during these interactions are often hard to trace back to specific causes or points of failure in the final result.
Therefore, in this work, we investigate an alternative approach ---
\textbf{how users' interaction behaviors\footnote{We defined the interaction behavior as the patterns of action event sequence (e.g., mouse click, scrolling, keypress). We refer to action event sequence as “action sequences” for brevity. } (mouse- and keyboard-based activities) correlate with their overreliance on the \ai}.
By capturing such behaviors, we can go beyond post-hoc analyses of final outputs and instead enable just-in-time detection of overreliance during task execution. This, in turn, opens the door to adaptive mitigation strategies (\eg \citet{10.1145/2380116.2380125}) that intervene only when necessary, without disrupting every user interaction.

Inspired by prior work that examines how crowdworkers’ interaction patterns relate to task quality on a HIT page~\cite{rzeszotarski2011instrumenting_the_crowd, gadiraju2019crowd, Han2016Crowdsourcing}, we investigate whether interaction behaviors similarly correlated to different levels of overreliance on \ai.
For example, imagine two users performing a writing task with ChatGPT. The \holSingle{}, after receiving the \llmresSingle{}, may superficially scan the content, copy and paste the whole text without carefully reviewing it. The \lolSingle{} on the other hand, may review the \llmresSingle{} thoroughly, and only selectively copy-paste task-relevant segments of sentences.

To identify user behaviors linked to overreliance, we first collected interaction data from 77 participants, who completed three distinct tasks by collaborating with a \ai pre-injected with plausible misinformation. Critically, this misinformation was designed to simulate naturally occurring LLM errors (\eg hallucinations, outdated information, and prompt injection attacks), while maintaining full experimental control over its content and placement.
To systematically measure user overreliance, we evaluated the extent to which participants’ final task answers were influenced by the pre-injected misinformation.
Existing researches~\cite{lai2019human, kim2024m, chan2024conversational, vasconcelos2023explanationsreduceoverrelianceai} have adopted this method as a balance between ecological plausibility and experimental reproducibility.
To reflect real-world scenarios, the tasks were tied to three of the most representative and frequently used information sources for accuracy judgment. Throughout the interaction, we logged a comprehensive set of action events, \eg click, scrolling, keypress. To analyze the behavior data, we employed a sequence-aware, state-of-the-art clustering method \cite{zhang2024mouse2vec}. To classify the behavioral patterns, we processed user action sequences, optimized clustering reliability, extracted typical behavioral patterns, and interpreted their cognitive implications.

As a result, our analyses identified five distinct behavioral patterns related to overreliance:
During task execution, \lol{} tended to carefully read task details at the beginning and work more independently after verifying the \llmresSingle{}. They navigated the interface in a fine-grained, context-specific manner, as they often modified text after pasting, rather than copying and pasting blindly.
In contrast, \hol{} frequently referred back to the \llmres{} throughout the task. Even when they initially hesitated on \llmresSingle{} and asked follow-up questions, they ultimately trusted and adopted it. Their navigation was coarser and visually guided, focusing on prominent or predictable interface areas-locations that were easy to locate---because they performed little editing and relied heavily on repeatedly copying entire sections at once.
We interpret participants' cognition by mapping each behavioral pattern to the cognitive theories whose core features closely align with the behavioral characteristics, and we reinforce these interpretations by cross-validating them with participants’ strategy reports.

Overall, we contribute a new understanding of overreliance on \ai by examining users' interaction behaviors.
Specifically:
\begin{itemize}[labelwidth=*,leftmargin=1.3em,align=left, topsep=0em]
    \item We collected a dataset~\footnote{https://github.com/CJunette/behavior\_indicator\_of\_overreliance} linking users' LLM interaction behaviors with quantitative measurements of their overreliance, which can serve as an entry point in understanding the relationship between overreliance and behavioral patterns.
    \item We adapted and refined a cluster-based analytical framework, tailoring it to quantify the relationship between behavioral data and overreliance---thereby enhancing the robustness of behavioral-to-reliance inference.
    \item We identified five distinct behavioral patterns corresponding to overreliance, complemented by empirical evidence, cognitive interpretation, and design implications for mitigating overreliance.
\end{itemize}
\section{Related Work}
We first review existing frameworks for \olm{} and limitations of current mitigations. Then we introduce existing outcome-oriented overreliance metrics and the feasibility of inferring states from interaction behaviors---highlighting the gap our work fills:
leveraging user behaviors to understand and characterize behavioral correlates of overreliance on conversational LLMs, as a first step toward real-time detection.

\subsection{Overreliance on AI: Existing Frameworks and Unique Challenges of \aiUp{}}
Previous works have defined overreliance on AI as
``users accepting incorrect AI recommendations''~\cite{passi2022overreliance},
``following its suggestions even when those suggestions are wrong and the person would have made a better choice on their own''~\cite{buccinca2021trustcognitiveforcing}.

Modern language models (\eg GPT~\cite{gpt4} and Gemini~\cite{gemini,gemini2}) primarily rely on the transformer architecture~\cite{vaswani2017attention} which learns from massive amounts of internet data to predict the next token (e.g, words in a sentence) based on how it is contextually related to the previous tokens.
While next-token prediction has shown a general power of solving various problems---from question answering to creative writing, the underlying transformer architecture necessarily brings along significant ``side effects''.
Generating unfaithful or misleading outputs (\ie hallucinations) is one of the key challenges in developing and using LLMs~\cite{bender2021dangers, brown2020language, chen2024combating,10.1145/3703155, 10.1145/3654777.3676366} while combating their ``misinformation harms''~\cite{weidinger2021ethicalsocialrisksharm, 10.1145/3654777.3676419,10.1145/3242587.3242666,10.1145/3654777.3676359} and aligning with human values~\cite{chen2023next,10.1145/3586183.3606763,10.1145/3654777.3676450}.
Specifically, misinformation harms include disseminating false or misleading information, causing material harm by disseminating false information (\eg in giving medical or legal advice), and leading users to perform unethical or illegal actions~\cite{weidinger2021ethicalsocialrisksharm, hendrycks2023aligningaisharedhuman}.

\aiUp{}'s unique characteristics, such as non deterministic outputs that confuse users and complicate verification~\cite{sanh2021multitask,arora2022ask}, erroneous backtracking when challenged~\cite{krishna2024understanding,laban2025llms}, sensitivity to indirect input attributes (e.g., epistemic markers, sycophancy, sandbagging)~\cite{zhou2023navigating,perez2023discovering,sharma2023towards}, and fast generation of high-volume novel content that imposes cognitive burden, raises verification costs, and leads users to mistake fluency/length for accuracy~\cite{metacognitive2024tankelevitch,topolinski2010immediate,ackerman2017meta}, fostering the risk of overreliance on AI.

Existing mitigations for overreliance on AI primarily include: explanations~\cite{goyal2023else, fok2024search, saunders2022self, si2023large}, uncertainty expressions~\cite{baan2023uncertainty, Kim2024, spatharioti2023comparing, vasconcelos2023generation}, providing recommendations only upon request~\cite{gaube2021ai, poursabzi2021manipulating}, altering speed of interaction~\cite{park2019slow}, and cognitive forcing functions~\cite{sarkar2024ai, danry2023don, buccinca2021trustcognitiveforcing}.
However, these mitigations have double-edged effects: explanations risk backfiring~\cite{si2023large, goyal2023else, steyvers2024calibration}, uncertainty expressions are plagued by poor model calibration~\cite{mielke2022reducing, radensky2023think}, and cognitive forcing functions risk under-reliance and impose additional cognitive burden~\cite{saunders2022self, danry2023don,perez2023towards}.
These shortcomings highlight that current one-size-fits-all mitigation strategies fail to account for the nuanced human-AI interaction.
It is necessary to develop adaptive mitigation \textit{that leverages real-time, process-level signals (e.g., behavioral traces) during the interaction process} ~\cite{bansal2021does, swaroop2024accuracy}.

\subsection{From Outcome to Process-Oriented Understandings of \olm{}}

Different from trust (i.e., a human attitude that may lead to reliance~\cite{howtoevaluate2021vereschak,hauptman2022componentstrust}), reliance itself refers to observable human actions, which can be operationalized via specific overreliance metrics and quantified.

Current research on measuring users' \olm{} mainly relies on \textit{outcome-oriented} metrics. Such studies infer overreliance levels from the final results of users’ task completion, with key measures including ``whether users adopt or agree with AI’s correct/incorrect outputs''~\cite{swaroop2025personalising,hunter2024monitoring,schemmer2022should}. For example, users are defined as overreliant if they make errors due to trusting LLMs’ hallucinations, or if they adopt AI suggestions at a high rate while disregarding independent judgment.
However, this outcome-oriented approach has two critical limitations:
First, it overlooks the \textit{process} (i.e., how humans interact with AI on the interface) that leads to outcomes, failing to capture behavioral details during interaction. As a result, it cannot distinguish between ``active rational reliance'' and ``passive blind compliance'', hindering the design of adaptive mitigations for different kinds of overreliance users.
Second, given the diversity of real-world LLM usage scenarios (e.g., academic writing, trip planning, programming), it is impractical to design scenario-specific outcome measures for every task, making it hard to establish a unified benchmark for \olm{} and undermining the generalizability of this approach.

While most research on measuring \olm{} focuses on outcome-oriented metrics, a smaller body has begun to explore other dimensions---specifically how users’ interaction behaviors and contextual features correlate with \olm{}. For instance, CUPS~\cite{mozannar2022reading} analyzes how programmers interact with code-recommendation systems, and proposes interaction inefficiencies and excessive time costs as potential behavioral signals of overreliance. Building on dual-process theories, prior work suggests that \olm{} often arises when users default to fast, intuitive, low-effort System~1 thinking instead of slower, effortful, analytical System~2 thinking~\cite{buccinca2021trustcognitiveforcing,doh2025beyond}, and that such cognitive strategies can be reflected in behavioral patterns, such as how thoroughly users read, edit, or verify AI outputs. Metacognitive research~\cite{metacognitive2024tankelevitch} links AI overreliance to users’ metacognitive challenges, which manifest as observable behaviors. REL-A.I.~\cite{zhou2024rel}, an interaction-centered framework, explores contextual features (e.g., knowledge domain, perceived LM competence) as contextual signals that modulate \olm{}, and relates these features to observable interaction patterns.
\citet{swaroop2024accuracy} investigated accuracy–time tradeoffs in AI-assisted decision making under time pressure, finding that overreliers and non-overreliers demonstrate differentiated use of AI assistance types, and that a user’s overreliance rate serves as a key behavioral correlate. However, existing studies on process-oriented correlates of \olm{} remain fragmented: they propose isolated signals but lack a systematic and quantified framework that organizes interaction behaviors into behavioral patterns that correlate with \olm{}.

\subsection{Interaction Behaviors as Performance Correlates}

Human behaviors are widely recognized as important predictors of individuals’ cognitive states and task outcomes, and user event logs---such as records of clicks, mouse movements, and keypresses---can effectively encode information about human interaction patterns and the subsequent task performance~\cite{understanding2022cao,marlow2015behavior,zhang2024mouse2vec}. \citet{jeffrey2011instrumenting} propose \textit{task fingerprinting}---a means to capture the process that crowdworkers use to complete a task, consisting of their interactions with the task interface, such as clicks, scrolls, and keypresses. Crowdspace supports the evaluation of crowdwork by combining workers’ behavioral traces with their task outputs through mixed-initiative machine learning, visualization, and interactive techniques~\cite{rzeszotarski2012crowdscape}. Other related works use behavioral traces to pre-select workers’ results~\cite{gadiraju2019crowd} and predict the quality of web page structure annotation~\cite{Han2016Crowdsourcing}.

In the era of Generative AI, this behavioral-inferential logic remains valid: \citet{ziegler2022productivity} identified a correlation between developer interactions and perceived productivity on GitHub through regression analysis, revealing that the acceptance rate of AI-generated suggestions is a more reliable predictor of productivity than alternative metrics.  CoAuthor captures rich interactions between writers and GPT-3 in creative writing tasks and demonstrated AI as a writing ``collaborator''~\cite{lee2022coauthor}.

All this prior work demonstrates the feasibility of mining users’ interaction behaviors to obtain process-level signals that correlate with high-level performance-related measures, although none has employed such an approach to understand \olm{}.
Our work addresses this gap by characterizing users’ interaction behaviors that correlate with outcome-level overreliance, providing a behavioral basis for future real-time detection and adaptive mitigation.
\section{Data Collection Study}
\label{sec:data_collection_experiment}

To investigate how users exhibit overreliance on \ai{} during interaction, we conducted a controlled laboratory experiment (approved by our institute’s IRB). To simulate diverse real-world use cases, we designed three task scenarios, each framed around a different information source. In every task, we deliberately injected plausible misinformation into the \llmres{}, simulating naturally occurring errors such as hallucinations, outdated knowledge or prompt injection. Users completed the tasks following a fixed procedure within a set time limit. Overreliance was defined as the extent to which users’ final submissions incorporated or were influenced by the injected misinformation. We collected detailed action logs of their interactions during the experiment. This setup allows us to systematically measure overreliance while also capturing the fine-grained behavioral patterns that accompany it.

\subsection{Study Design}

\paragraph{Participants}
We recruited \participantNum{} participants from two local universities.
Participants averaged 22.40 years old (SD=9.22), with 30 having high school/some college, 31 a Bachelor’s, 10 a Master’s, and 2 a Doctoral degree.
On a 5-point Likert scale, participants' familiarity with \ai averaged 3.0 (SD = 1.2) out of 5, and participants' frequency of \ai usage averaged 3.8 (SD = 1.8) out of 5.
The experiment took approximately 1.5 hours, and participants received \$25 as compensation.

\paragraph{Procedure}
\label{subsec:procedure}
Prior to the experiment, participants provided informed consent (Appendix~\ref{subsubsec:informedConsentStatement}), completed a demographic and background survey, and were guided via a video tutorial to install a browser extension---which collected behavioral data throughout task completion.
After each task, participants submitted the answer and uploaded behavioral logs from the extension. Upon completing all tasks, they filled out a questionnaire assessing LLM trust, task familiarity, and decision-making strategies (Section \ref{subsec:questionaire_after_task}).
Technical assistance was available via online chat throughout the experiment. Total duration was approximately 1.5 hours.

\begin{figure}
    \centering
    \Description{This figure displays a computer screen split into two vertical halves, showing the interface participants used in an experiment.
    Left Half: Titled "H2B Summary Task," this side shows a task-related page. At the top, there is a title "Task Begin" and a question, "Why do some people sneeze so loudly?". Below the question is a picture of a man in profile, outdoors in what appears to be cold weather, sneezing forcefully into the air. Below the image is a block of text that begins, "When I sneeze, everyone knows about it." The text is an article for the user to read or summarize.
    Right Half: Titled "NextChat," this side displays a Large Language Model (LLM) chat interface. The main chat window is empty with a `"New Conversation" heading. At the bottom is a text input box that says, "Ctrl + Enter to send, / to search prompts, : to use commands," with a "Send" button to the right. The overall impression is that a user is expected to use the LLM on the right to help with the task presented on the left.
    The caption at the bottom reads: "Fig. 1. The Interface Setup in Experiment. Participants are asked to use a split-screen, the left half of the screen will display the task-related page, and the right half will display the LLM page as well as the page for search engine."}
    \includegraphics[width=1\linewidth]{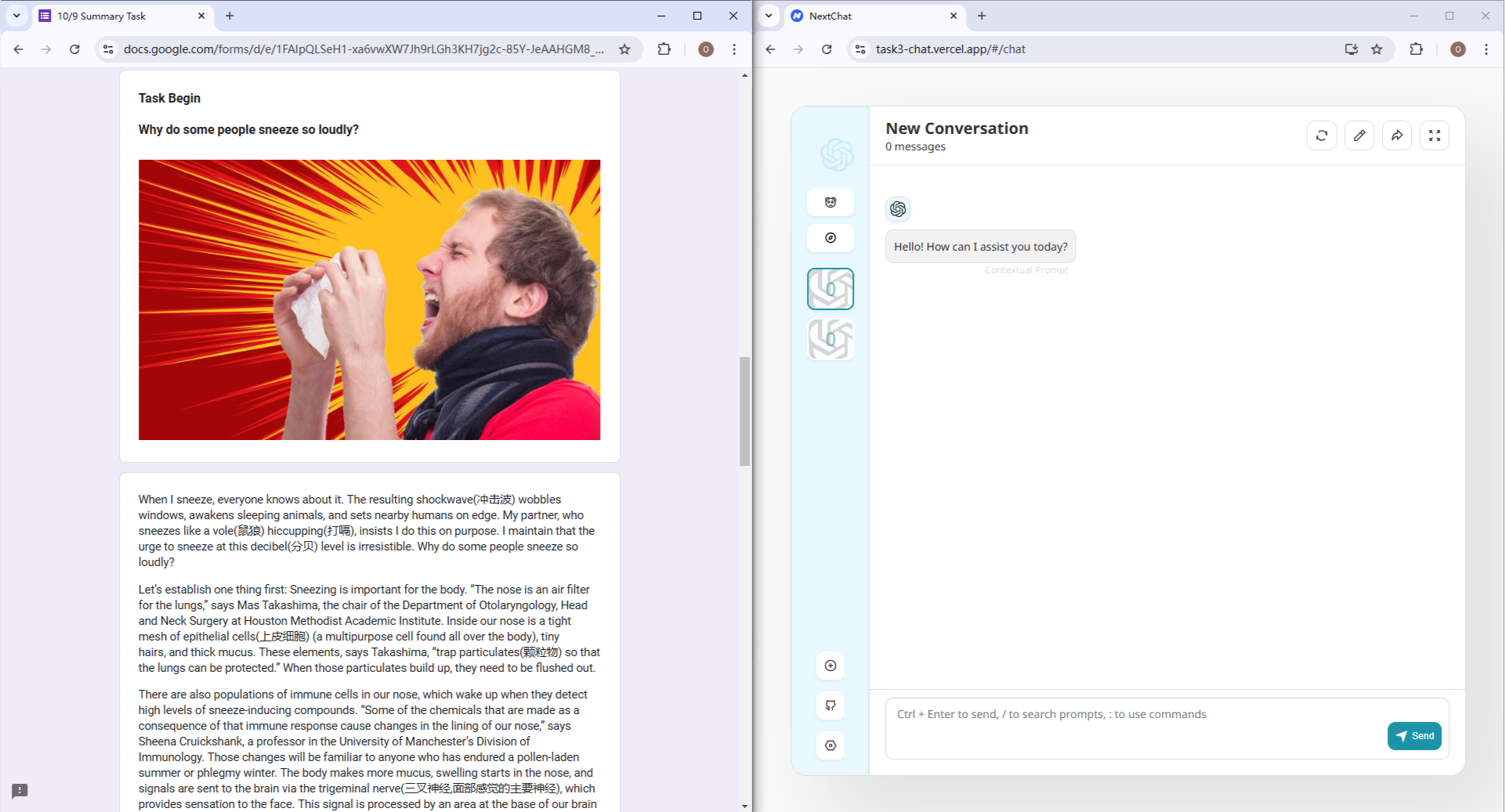}
    \caption{Interface setup in the experiment. Participants are asked to use a split-screen, the left half of the screen will display the task-related page, and the right half will display the LLM page as well as the page for search engine.}
    \label{fig:split_screen}
\end{figure}

\paragraph{Task Platform}
We conducted the experiment online, where participants opened the experiment link in a Chrome web browser on their own PCs to participate. Based on the open-source project ChatGPT-Next-Web\footnote{\href{https://github.com/ChatGPTNextWeb/ChatGPT-Next-Web}{https://github.com/ChatGPTNextWeb/ChatGPT-Next-Web}} and the GPT-4 model, we developed a privatized dialogue platform for our experiment, which serves as the \ai.
Figure~\ref{fig:split_screen} demonstrates the setup participants use for the user study.

\paragraph{Behavioral Data Logging}
\label{subsec:data_collection}
We used Chrome extensions to collect user behavior data, via JavaScript scripts that utilize event listeners to monitor user actions on the web page, including \textit{mouseMovement}, \textit{clicks}, \textit{scrolling}, \textit{keypress}, \textit{copy}, \textit{paste}, \textit{highlighting}, \textit{delete}, and \textit{idleness} (defined as 3 seconds of inactivity). Additionally, we monitored window switch behaviors on the LLM interface, which indicate that users are using alternative tools other than the LLM.

We logged the user actions on task page and LLM page separately into two files.

Please refer to Table \ref{tab:action_and_action_featuers_1} for details of all the actions and  features we collected.

\begin{table*}[ht]
    \fontsize{8}{9}\selectfont  
    \renewcommand{\arraystretch}{1.3}  
    \centering
    
    \begin{tabular}{
        >{\raggedleft\arraybackslash}p{0.15\textwidth}  
        >{\raggedright\arraybackslash}p{0.25\textwidth}  
        >{\raggedleft\arraybackslash}p{0.2\textwidth}  
        >{\raggedright\arraybackslash}p{0.3\textwidth}  
        }
        \toprule  
        \textbf{Action}                       & \textbf{Action Description}                                                                & \textbf{Action Attribute}          & \textbf{Action Feature Description}    \\
        \midrule  
        
        \multirow{2}{*}{mouseMovement}        & \multirow{2}{*}{When the mouse moves.}                                                     & \texttt{total\_mouse\_movement}    & The total distance of mouse movement.  \\  
                                              &                                                                                            & \texttt{mouse\_movement\_duration} & The duration of a mouse move.          \\
        \midrule
        
        click                                 & When the mouse clicks.                                                                     & -                                  & -                                      \\
        \midrule
        
        \multirow{3}{*}{scroll \& mousewheel} & Mouse wheel actions in scrolling. Scroll event triggers when an element has been scrolled. & \texttt{scroll\_duration}          & The duration of a scroll.              \\
        \cmidrule(lr){3-4}  
                                              &                                                                                            & \texttt{scroll\_distance}          & The distance of a scroll.              \\  
        \cmidrule(lr){3-4}
                                              &                                                                                            & \texttt{scroll\_direction}         & The direction of a scroll.             \\
        \midrule
        
        \multirow{2}{*}{keypress}             & \multirow{2}{0.25\textwidth}{When the user presses keys                                                                                                                  \\ on the keyboard.} & \texttt{keypress\_duration} & The duration of a keypress. \\
        \cmidrule(lr){3-4}
                                              &                                                                                            & \texttt{keypress\_keyCount}        & The count of inputs during a keypress. \\
        \midrule
        
        copy                                  & When the user copies text.                                                                 & \texttt{copy\_textLength}          & The length of copied text.             \\
        \midrule
        
        paste                                 & When the user pastes text.                                                                 & \texttt{paste\_textLength}         & The length of pasted text.             \\
        \midrule
        
        highlight                             & When the user selects certain text.                                                        & \texttt{highlight\_textLength}     & The length of highlighted text.        \\  
        \midrule
        
        \multirow{2}{*}{delete}               & When the user deletes text using actions such as backspace, delete, or cut (Ctrl+X).       & \texttt{delete\_duration}          & The duration of a deletion.            \\  
        \cmidrule(lr){3-4}
                                              &                                                                                            & \texttt{delete\_keyCount}          & The count of deletions.                \\  
        \midrule
        
        idleness                              & When the user doesn't input any action for over 3 seconds.                                 & \texttt{idle\_duration}            & The duration of an idleness.           \\
        \midrule
        
        * element switch                      & When the user conducts an element switch.                                                  & -                                  & -                                      \\  
        \midrule
        
        * tab switch                          & When the user switches to another tab within the LLM window.                               & -                                  & -                                      \\
        \midrule
        * prompt input                        & When the user writes prompts within the LLM window.                                        & -                                  & -                                      \\
        \bottomrule
    \end{tabular}
    \caption{User Behaviors and Corresponding Features Collected for the Study. This table presents the preprocessed user behaviors and their associated attributes/features collected via Chrome extensions (in both LLM and Task pages) using JavaScript event listeners. The collected behaviors include both general web actions (e.g., mouse movement, clicks, scrolling, and keypresses) and LLM page-specific actions (marked with *: element switch, tab switch within the LLM window, and prompt input in the LLM window). For each behavior, the table details its description, action attributes, and feature descriptions.}
    \label{tab:action_and_action_featuers_1}
    \Description{This is a table with four columns: "Action," "Action Description," "Action Attribute," and "Action Feature Description." It details the user behaviors that were tracked during the study and the specific data points (features) collected for each behavior.
    mouseMovement: When the mouse moves. The feature collected is total_mouse_movement_duration, described as the total distance of mouse movement.
    click: When the user clicks. No specific attribute is listed.
    scroll \& mousewheel: Mouse wheel actions in scrolling.
    scroll_duration: The duration of a scroll.
    scroll_distance: The distance of a scroll.
    scroll_direction: The direction of a scroll.
    keypress: When the user presses keys on the keyboard.
    keypress_duration: The duration of a keypress.
    keypress_keyCount: The count of inputs during a keypress.
    copy: When the user copies text. The feature is copy_textLength, the length of copied text.
    paste: When the user pastes text. The feature is paste_textLength, the length of pasted text.
    highlight: When the user selects certain text. The feature is highlight_textLength, the length of highlighted text.
    delete: When the user deletes text (e.g., backspace, delete, cut).
    delete_duration: The duration of a deletion.
    delete_keyCount: The count of deletions.
    idleness: When the user doesn't input any action for over 3 seconds. The feature is idle_duration, the duration of an idleness.
    element switch: When the user conducts an element switch. No attribute listed.
    tab switch: When the user switches to another tab within the LLM window. No attribute listed.
    prompt input: When the user writes prompts within the LLM window. No attribute listed.
    The caption notes that behaviors marked with an asterisk (*) are page-specific actions.
    }
\end{table*}

\subsection{Task Design and Misinformation Injections}
\label{subsec:design}

To determine the task design, we conducted expert interviews with LLM misinformation researchers to discuss and justify the guiding principles for task selection.

\begin{enumerate}
    \item The task should be commonly recognized as tasks where an LLM is typically used.
    \item The tasks cover three common sources to judge LLM-provided information: \one judging by personal knowledge (common sense), \two judging by limited contextual information, and \three judging by external information, e.g., via online search.
    \item The tasks should have clear measurable outcomes, allowing a straightforward assessment of overreliance.
\end{enumerate}

As a result, we chose three tasks---\expOne{}, \expTwo{}, and \expThree{}, which we describe in Table \ref{tab:task_overview}. Check Appendix \ref{appendix:task_details} for details about each task.
\expOne{} included two trials under two conditions (with \vs w/o LLM), yielding $3 \times 2 \times 2=12$ different task orders, which we counterbalanced across participants.

Pilot studies were first used to determine the appropriate time allocation based on the difficulty of Task 1, ensuring that participants could complete the task with the assistance of the LLM. Using this time constraint as a reference, we subsequently calibrated the difficulty of Task 2 and Task 3. For Task 2, we selected an essay of moderate difficulty, and for Task 3, we adjusted the required number of submissions. These adjustments ensured that participants would need LLM assistance to complete the tasks within the allotted time.
Furthermore, before beginning the tasks, participants were instructed to complete the task with the LLM and to engage seriously with the experiment; otherwise, their compensation might be reduced.

\begin{table*}[ht]
    \renewcommand{\arraystretch}{1.6}
    \centering
    \begin{tabular}{
        >{\raggedleft\arraybackslash\small}p{0.06\textwidth}
        >{\raggedright\arraybackslash\small}p{0.13\textwidth}
        >{\raggedright\arraybackslash\small}p{0.12\textwidth}
        >{\raggedright\arraybackslash\small}p{0.32\textwidth}
        >{\raggedright\arraybackslash}m{0.22\textwidth}
        }
        \toprule
        \textbf{Task}         & \textbf{Description}                                    & \textbf{Design}                               & \textbf{How LLM Generates Misinfo.}                                                                                                                                                 & \textbf{Overreliance Metrics}                                                      \\
        \midrule

        Quiz Solving          & Rank the importance of 15 items in a survival scenario. & 2 trials × \{w/, w/o\} LLM, 15 mins per trial & Manually create an incorrect ranking and ask the LLM to provide a reasonable explanation (e.g., "Matches are crucial for survival on the moon due to limited oxygen availability"). & $\displaystyle \frac{(S_{\text{w/LLM}} - S_{\text{w/o LLM}}) - \min}{\max - \min}$ \\
        \midrule

        Article Summarization & Summarize a 700-word scientific article.                & Single task, with LLM, 15 mins                & Ask the LLM to generate a summary with errors (e.g., changing "sneezing is to flush out particulates" in the original text to "sneezing is to flush out cells").                    & $\displaystyle \frac{\text{Retained misinfo. count}}{\text{Total misinfo. count}}$                          \\
        \midrule

        Trip Planning         & Collect information for trip planning.                  & Single task, with LLM, 15 mins                & Ask the LLM to generate suggestions with errors (e.g., generating locations that are not in the destination, or providing incorrect admission prices).                              & $\displaystyle \frac{\text{Retained misinfo. count}}{\text{Total misinfo. count}}$    \\
        \bottomrule
    \end{tabular}

    \caption{Details of the Three Tasks Designed for the Study. The table includes each task's identifier, description, design, misinformation injection method, and metric for measuring user overreliance on LLMs.}
    \Description{
    This table has five columns: "Task," "Description," "Design," "How LLM Generates Misinfo.," and "Overreliance Metrics." It outlines the three experimental tasks.
    Task 1: Quiz Solving
    Description: Rank the importance of 15 items in a survival scenario.
    Design: 2 trials × {w/, w/o} LLM, 15 mins per trial.
    How LLM Generates Misinfo.: Manually create a wrong ranking and ask the LLM to provide a reasonable explanation (e.g., "Matches are crucial for survival on the moon due to limited oxygen availability").
    Overreliance Metrics: A formula is provided: (S<sub>w/LLM</sub> − S<sub>w/o LLM</sub>) − min / max − min.
    Task 2: Article Summarization
    Description: Summarize a 700-word scientific article.
    Design: Single task, with LLM, 15 mins.
    How LLM Generates Misinfo.: Ask the LLM to generate a summary with errors (e.g., changing "sneezing is to flush out particulates" in the original text to "sneezing is to flush out cells").
    Overreliance Metrics: Retained misinfo. count / Total misinfo. count.
    Task 3: Trip Planning
    Description: Collect information for trip planning.
    Design: Single task, with LLM, 15 mins.
    How LLM Generates Misinfo.: Ask the LLM to generate suggestions with errors (e.g., generating locations that are not in the destination, or providing incorrect admission prices).
    Overreliance Metrics: Retained misinfo. count / Total misinfo. count.
    }
    \label{tab:task_overview}
\end{table*}

To systematically study overreliance, we injected task-specific plausible misinformation into LLM outputs. Following prior work \cite{lai2019human, kim2024m, chan2024conversational, vasconcelos2023explanationsreduceoverrelianceai}, this design simulates natural misinformation while enabling systematic evaluation of overreliance by measuring misinformation in participants' final outputs.
In the rest of this section, we describe the detailed design for each task, including: 1) the rationale for choosing the task, 2) the nature of misinformation injected, and 3) the evaluation of overreliance, following the framework below.

\subsubsection{\expOneCaps}

This task was selected to simulate real-world scenarios where users rely on LLMs for decision-making, contrasting behaviors when participants use only personal common sense versus LLM-provided information. Specifically, participants ranked the importance of survival items in two scenarios (survival on the moon and survival in the desert) and under two conditions (\withAI{} and \withoutAI{}). The four conditions were counterbalanced to mitigate learning effects.

\textbf{Misinformation Injection.}
To mimic real-world LLM hallucinations while avoiding obvious absurdity, we altered the expert-validated ranking of survival items to generate plausible yet incorrect orderings. The LLM was instructed to consistently output this erroneous ranking and provide convincing but false justifications when participants requested explanations.
\textbf{Evaluation Metrics.}
We used the NASA standard scoring algorithm (Appendix A.3) to calculate a deviation score (S), where lower scores indicate closer alignment with the ground truth \cite{larson2019team, nowak2023hear, esfandiari2024meet}.
Overreliance was quantified as the difference in performance scores between the \withAI{} and \withoutAI{} conditions. The raw difference was normalized using the minimum and maximum scores across participants.

\subsubsection{\expTwoCaps}

This task reflects practical use cases of LLMs for text processing (e.g., literature abstracting, news summarization). It assesses participants’ ability to evaluate whether LLMs’ interpretations align with the content provided by users.

\textbf{Misinformation Injection.} According to previous methods \cite{lai2019human, kim2024m, chan2024conversational, vasconcelos2023explanationsreduceoverrelianceai}, we generate detail-tampered misinformation by instructing the LLM to slightly alter facts from the original text. Eleven misinformation instances requiring contextual reasoning to detect were manually selected and embedded in the LLM’s system prompt; the LLM was instructed to persist with these errors even if challenged.

\textbf{Evaluation Metrics.} Normalizing the count of retained misinformation against the total number of injected instances. Overreliance is calculated as the occurrence of misinformation in users' summary answers.

\subsubsection{\expThreeCaps}

This task simulates real-world scenarios where users rely on LLMs for accessing time-sensitive information (e.g., API integration, trip planning)---contexts in which external search engines serve as the primary tool to verify such time-sensitive content. Specifically, participants were instructed to plan a trip to Copenhagen for a specific date.

\textbf{Misinformation Injection.}
Similar to \expTwoCaps, we first asked the LLM to generate accurate travel recommendations for Copenhagen, then instructed it to modify these recommendations into fact-checkable misinformation. Twenty such verifiable misinformation instances were selected. The LLM was further instructed to provide these pieces of misinformation with plausible justifications whenever participants inquired about relevant topics.

\textbf{Evaluation Metrics.} Instances of misinformation retained in participants’ itineraries. Given the task’s openness, GPT-4 was used for automated detection via named entity recognition and logical inference (e.g., identifying false admission prices or fabricated attraction locations).
The evaluation normalizes the count of retained misinformation against the total injected instances.
\section{Analysis}
\label{sec:analysis_method}

    In this section, we present the methods used to process and analyze the collected interaction data. The goal is to uncover behavioral patterns correlated with user \olm{}.
    We processed the interaction logs to enable meaningful behavioral analysis. Preprocessing ensured data completeness and interpretability. Vectorization converted events into a format suitable for computational analysis. Segmentation allowed examination of behavior at different temporal scales. Autoencoder embedding produced consistent low-dimensional representations of variable-length sequences. Clustering identified recurring behavioral patterns, and post-clustering selection retained only robust, informative clusters for further interpretation of users’ cognitive and behavioral strategies.

\begin{figure*}[t]
    \centering
    \includegraphics[width=\linewidth]{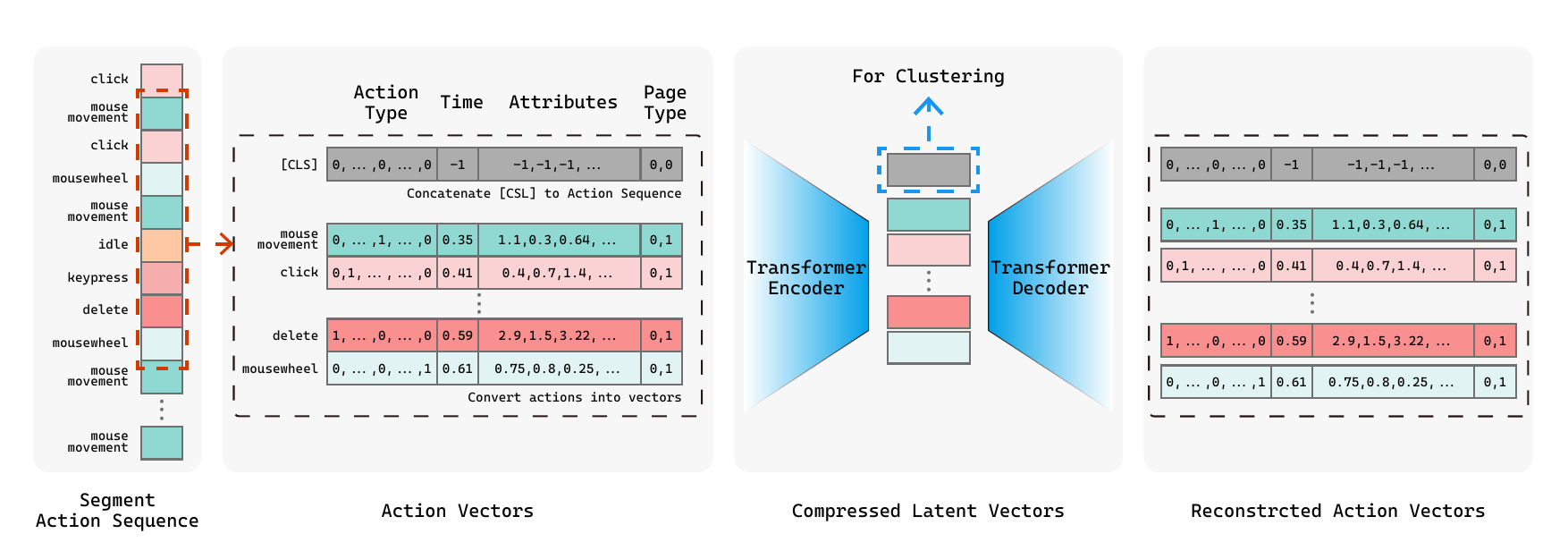}
    \caption{\textbf{Overview of the Analysis Pipeline.} We segment user interaction logs into overlapping time-based windows, encode each sequence into standardized feature vectors, use an autoencoder to produce compact sequence embeddings, and cluster these embeddings to identify recurring behavioral patterns. Selected clusters are interpreted in terms of user overreliance.}
    \label{fig:auto_encoder}
    \Description{
    This is a flowchart diagram illustrating the data analysis process, moving from left to right.
    1. Segment Action Sequence: On the far left, a vertical bar represents a sequence of user actions like "click," "mouse movement," "click," etc. A dashed red box highlights a segment of this sequence.
    2. Action Vectors: An arrow points from the segmented sequence to a representation of "Action Vectors." This shows the user actions converted into a numerical table (vectors) with columns for "Action Type," "Time," and "Attributes."
    3. For Clustering (Autoencoder): An arrow points to the central part of the diagram, which depicts a transformer autoencoder. The action vectors are fed into a "Transformer Encoder," which compresses them into "Compressed Latent Vectors" (shown as smaller, abstract shapes). These are then fed into a "Transformer Decoder" to create "Reconstructed Action Vectors." An arrow labeled "For Clustering" points up from the compressed latent vectors.
    4. Reconstructed Action Vectors: On the far right, the reconstructed vectors are shown in the same table format as the initial action vectors, representing the output of the autoencoder.
    The caption reads: "Fig. 2. Overview of the Analysis Pipeline. We segment user interaction logs into overlapping time-based windows, encode each sequence into standardized feature vectors, use an autoencoder to produce compact sequence embeddings, and cluster these embeddings to identify recurring behavior patterns. Selected clusters are interpreted in terms of user overreliance."
    }
\end{figure*}

    \paragraph{Preprocessing}
    \label{subsec:preprocessing}
    We first preprocessed the data for downstream analysis. Incomplete logs resulting from upload errors or user-interaction issues were removed. To reduce redundancy and improve interpretability, low-level events (e.g., mouse movements, scrolls, keystrokes) were merged into higher-level action events using temporal and semantic heuristics. All events were then labeled by page context (Task or LLM), merged by timestamp, temporally aligned, and normalized so that timestamps range from 0 to 1, representing the full session duration. Details of filtering, merging, and normalization are provided in Appendix~\ref{appendix:preprocessing}.

\paragraph{Vectorization}
\label{subsubsec:vectorization}
To ensure that log events could be consistently encoded and processed, we vectorized all event data.
Each action event is encoded as a 37-dimensional feature vector representing four components:
the action type (15 dimensions), timestamp (1 dimension), page type (2 dimensions), and various type-specific continuous and categorical attributes (19 dimensions). Table~\ref{tab:action_feature_encoding} details each feature and the number of dimensions allocated to it (Appendix~\ref{appendix:vector_dimensions}). These features cover all parameters recorded in the event log (i.e., the Action Attributes listed in Table~\ref{tab:action_and_action_featuers_1}), ensuring that our event encoding preserves the full information present in the original logs.

\paragraph{Segmentation}
\label{subsubsec:segmentation}

To analyze user behavior at varying temporal granularities, we segmented each session into fixed-duration windows ranging from 10 to 60 seconds, with a 1-second stride. The rationale behind this choice is that shorter windows yield isolated action events that lack meaningful patterns, whereas longer windows often produce highly repetitive sequences. These segments serve as the fundamental behavioral units for encoding and clustering. Segments inherit the overreliance label (one score per task per participant), allowing us to examine how localized behaviors relate to overall task performance.

\paragraph{Autoencoder Embedding}
\label{subsubsec:auto_encoder_model}
To produce consistent representations across variable-length action sequences, we used a transformer-based autoencoder that projects each sequence into a low-dimensional latent space while preserving essential behavioral information.

A special \texttt{[CLS]} token was prepended to each input sequence and served as a global summarization vector. The encoder maps the entire sequence into latent space, and the decoder seeks to reconstruct the original. The final embeddings used for clustering are taken from the \texttt{[CLS]} output vector of the encoder.

We trained a separate model for each combination of task and window size (3 tasks × 6 windows = 18 models). A full model specification and training loss breakdown are provided in Appendix~\ref{appendix:autoencoder_architecture}.

\paragraph{Clustering}
\label{subsubsec:clustering}

We clustered the \texttt{[CLS]} latent vectors using DBSCAN~\cite{ester1996dbscan}. To ensure robustness, we explored multiple clustering thresholds and neighborhood sizes (eps and min\_samples), retaining only those clusters that appeared consistently across parameterizations.

For the test set, we assigned cluster labels using k-nearest neighbors (k = 5) in the embedding space. Clustering and assignment parameters are described in Appendix~\ref{appendix:clustering_details}.

\paragraph{Post-Clustering Selection}
\label{subsubsec:post_clustering_process}

To retain only meaningful and generalizable patterns, we applied two criteria to filter clusters:

\begin{itemize}
    \item \textbf{Intrinsic Similarity}: The distribution of overreliance scores in the training and test set assigned to a cluster must not significantly differ (two-tailed t-test, $p > 0.05$). It indicates that members within a cluster exhibit a similar distribution of overreliance.
    \item \textbf{Predictive Capability}: The average overreliance score of training members must accurately predict that of test members (details in Appendix~\ref{appendix:computing_predictive_score_for_test}). It indicates that the cluster’s average overreliance level reliably reflects its members’ tendencies.

\end{itemize}

For each selected cluster, we extracted the 20 latent vectors nearest the cluster centroid and retrieved their corresponding action sequences. These representative sequences were manually analyzed to interpret behavioral and cognitive strategies exhibited by users under varying levels of overreliance. Overreliance levels were binned into high, neutral, and low categories for interpretive purposes as described in Appendix~\ref{appendix:overreliance_levels}.

\section{Findings}
\label{sec:quantitative_findings}

\begin{figure*}[htbp]
    \centering
    \begin{subfigure}[b]{0.32\textwidth}
        \includegraphics[width=\textwidth]{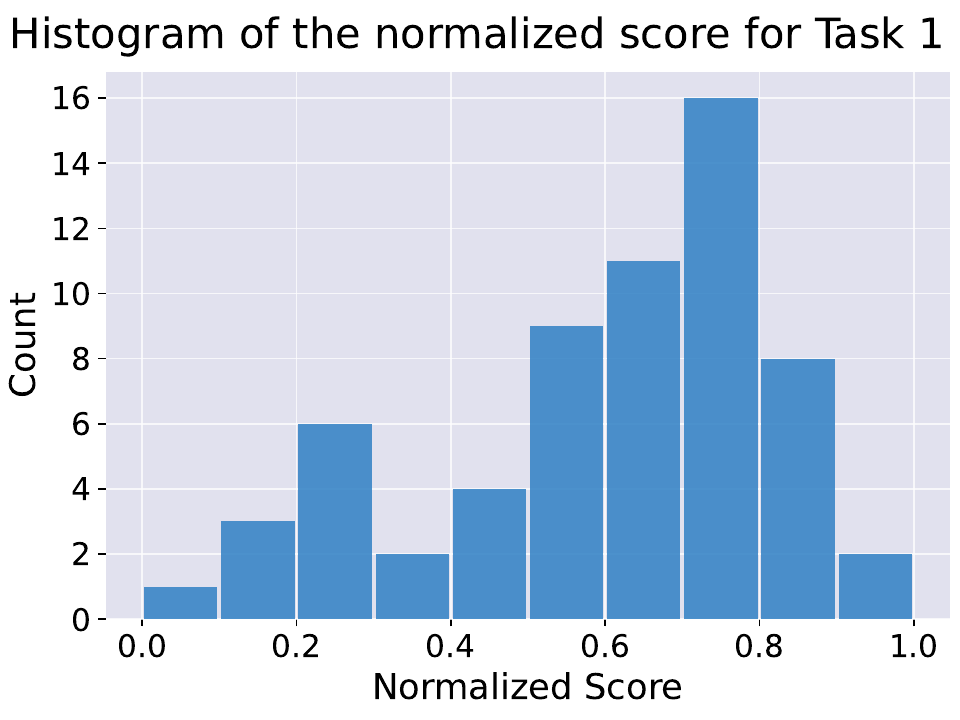}
        \caption{\expOne{}}
        \label{fig:histogram_of_normalized_score_task_1}
    \end{subfigure}
    \hfill
    \begin{subfigure}[b]{0.32\textwidth}
        \includegraphics[width=\textwidth]{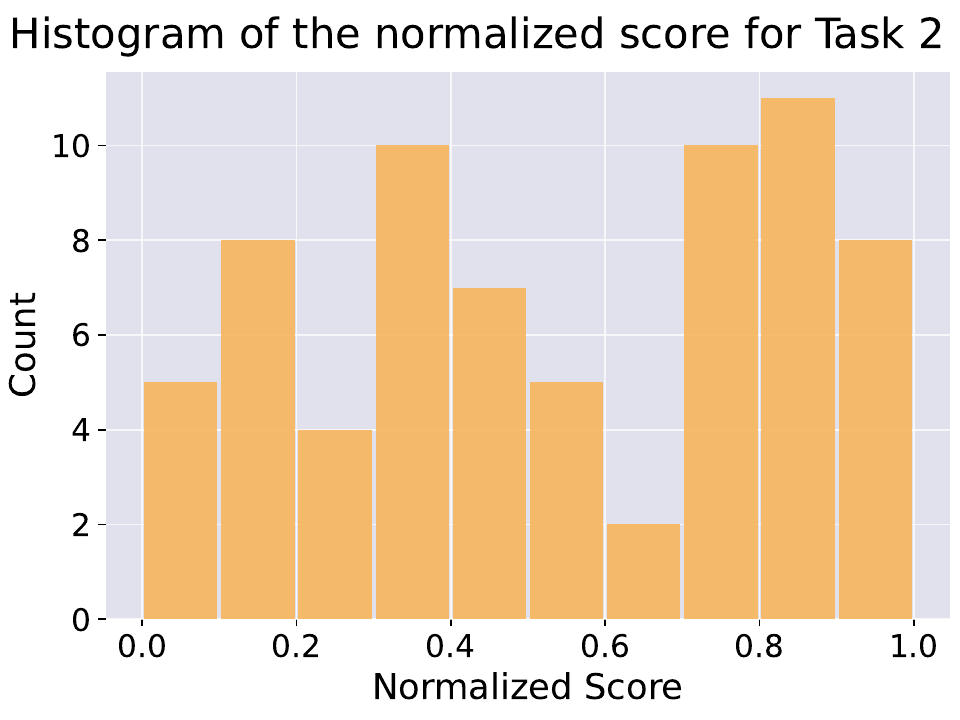}
        \caption{\expTwo{}}
        \label{fig:histogram_of_normalized_score_task_2}
    \end{subfigure}
    \hfill
    \begin{subfigure}[b]{0.32\textwidth}
        \includegraphics[width=\textwidth]{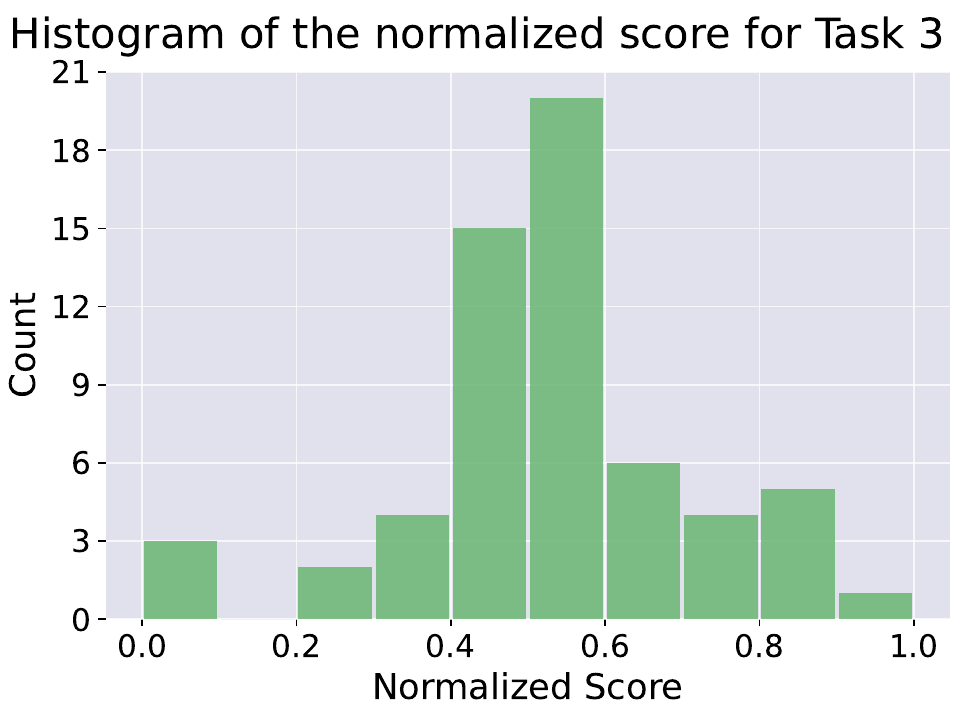}
        \caption{\expThree{}}
        \label{fig:histogram_of_normalized_score_task_3}
    \end{subfigure}
    \caption{Histograms of normalized overreliance scores for three distinct tasks, where each subfigure corresponds to one task, the vertical axis (\textit{Count}) represents the number of participants falling into each score bin, and the horizontal axis (\textit{Normalized Score}) represents the normalized measure of overreliance on AI. A lower normalized score indicates a lower degree of overreliance on AI, while a higher score signifies a greater level of overreliance. The three tasks include: (a) \expOne{}, (b) \expTwo{}, (c) \expThree{}.}
    \Description{
    This figure contains three histograms, labeled (a), (b), and (c), each representing a different task from the study. The horizontal axis for all three is "Normalized Score" ranging from 0.0 to 1.0, and the vertical axis is "Count," representing the number of participants. A lower score indicates less overreliance on AI, while a higher score indicates greater overreliance.
    (a) quiz solving: This histogram is colored blue. The scores are distributed across the range, with the highest peak (modal bin) between 0.6 and 0.7, where nearly 16 participants fall. The distribution is roughly bell-shaped but slightly skewed to the left.
    (b) article summarization: This histogram is colored orange. The distribution has two main peaks, one around 0.3-0.4 and another higher peak around 0.7-0.8, with about 11 participants in that bin. This suggests a bimodal distribution of participants' overreliance.
    (c) trip planning: This histogram is colored green. The distribution is heavily skewed to the left, with the vast majority of participants (nearly 21 in the highest bin) scoring between 0.4 and 0.6. Very few participants have very high or very low scores.
    }
    \label{fig:histogram_of_normalized_score}
\end{figure*}

Figure \ref{fig:histogram_of_normalized_score} shows the histogram of normalized overreliance scores
for three different tasks.
A lower overreliance score indicates a lower degree of overreliance on AI, while a higher value signifies a greater level of overreliance.

As mentioned in Section \ref{subsubsec:metric_exp_one}, it is essential to ensure no significant difference in $\text{score}_\text{trial}$ between the two trials of Task 1. Therefore, we conducted further validation on the collected data.

After clustering the 18 combinations, we identified a total of 154 distinct clusters. Applying the selection criteria from Section

\ref{sec:analysis_method}, we filtered these down to
54 core valid clusters with good predictive capability and strong overreliance similarity.
Among these, 18 salient clusters have significantly high or low overreliance in the training set, and 36 clusters correspond to cases with no significant difference.
The presence of these clusters suggests that our method can distinguish behavioral patterns associated with higher versus lower overreliance levels in our dataset.
We also retrieved participants’ answers to the decision-making strategies in post-task questionnaire after each task. We first linked each behavioral cluster to its corresponding participants and tasks, and then retrieved the post-task strategy reports for those cluster members. We coded these reports using a codebook aligned with our identified behavioral patterns and theoretical constructs related to \olm{}, iteratively refining codes and themes in team meetings and writing analytic memos to capture how self-reported strategies supported or challenged our behavioral interpretations.

\newcommand{\categoryOneWidth}{5.5em}
\newcommand{\categoryTwoWidth}{8em}
\newcommand{\distributionWidth}{10em}
\newcommand{\overrelianceWidth}{5.5em}
\newcommand{\actionSequenceWidth}{12em}
\newcommand{\interpretedBehaviorWidth}{12em}
\newcommand{\inferredCognitivePatternWidth}{12em}

\begin{table*}[htbp]
    \centering
    \small
    \renewcommand{\arraystretch}{1.3}
        \begin{tabular}{
            >{\raggedleft\arraybackslash}m{0.12\textwidth}
            >{\raggedright\arraybackslash}m{0.70\textwidth}
            }
            \toprule
            \cellcolor[HTML]{C0C0C0}\textbf{Overreliance} & \cellcolor[HTML]{C0C0C0}\textbf{Action Sequences, Interpreted Behavior, Cognitive Process}                                                                                                                                              \\
            \midrule
            \multicolumn{2}{c}{\cellcolor[HTML]{EFEFEF}\textbf{(a) Frequency of Copying-Pasting}}                                                                                                                                                                                                   \\
            \multicolumn{2}{c}{\cellcolor[HTML]{EFEFEF}Task: Trip planning, Window: 10–60s, $N=32$}                                                                                                                                                                                                 \\
            \cmidrule(lr){1-2}
            \cellcolor[HTML]{FFFFFF}High                  & \cellcolor[HTML]{FFFFFF}Repeated \textit{copy-paste}, no text editing. \textcolor{brown}{Adopt \llmres{} directly.}                                                                                                                     \\
            \cellcolor[HTML]{FFFFFF}Low                   & \cellcolor[HTML]{FFFFFF}\textit{Keypress/delete} before/after pasting, less \textit{copy-paste}. \textcolor{brown}{Cautious copying + post-paste editing.}                                                                              \\
            \midrule
            \multicolumn{2}{c}{\cellcolor[HTML]{EFEFEF}\textbf{(b) Focused Task Comprehension at the Start}}                                                                                                                                                                                        \\
            \multicolumn{2}{c}{\cellcolor[HTML]{EFEFEF}Task: Article summarization (Window: 10–50s, $N=22$); Trip planning (Window: 10–50s, $N=13$)}                                                                                                                                                \\
            \cmidrule(lr){1-2}
            \cellcolor[HTML]{FFFFFF}Low                   & \cellcolor[HTML]{FFFFFF}Consecutive alternating \textit{mousewheel + scroll} at task start. \textcolor{brown}{Carefully read task initially.}                                                                                           \\
            \midrule
            \multicolumn{2}{c}{\cellcolor[HTML]{EFEFEF}\textbf{(c) Frequency of Referring LLM’s responses}}                                                                                                                                                                                         \\
            \multicolumn{2}{c}{\cellcolor[HTML]{EFEFEF}Task: Quiz solving, Window: 10s, $N=25$}                                                                                                                                                                                                     \\
            \cmidrule(lr){1-2}
            \cellcolor[HTML]{FFFFFF}High                  & \cellcolor[HTML]{FFFFFF}Alternated between LLM \& Task pages, single \textit{mousewheel + mouseMovement} on LLM then \textit{click} on Task each time. \textcolor{brown}{Refer to \llmres{} per ranking.}                               \\
            \cellcolor[HTML]{FFFFFF}Low                   & \cellcolor[HTML]{FFFFFF}Scroll (\textit{mousewheel}) on LLM page, then switched to Task page \& do \textit{mouseMovement + click} series. \textcolor{brown}{After referring to \llmresSingle{}, finish all the rankings independently.} \\
            \midrule
            \multicolumn{2}{c}{\cellcolor[HTML]{EFEFEF}\textbf{(d) Coarse- vs. Fine-Grained Locating \& Editing}}                                                                                                                                                                                   \\
            \multicolumn{2}{c}{\cellcolor[HTML]{EFEFEF}Task: Article summarization, Window: 50–60s, $N=34$}                                                                                                                                                                                         \\
            \cmidrule(lr){1-2}
            \cellcolor[HTML]{FFFFFF}High                  & \cellcolor[HTML]{FFFFFF}Consecutive \textit{mousewheel/scroll + mouseMovement}, lengthy \textit{copy-paste} at end. \textcolor{brown}{“Rough” navigation (prominent areas).}                                                            \\
            \cellcolor[HTML]{FFFFFF}Low                   & \cellcolor[HTML]{FFFFFF}Few \textit{mousewheel/scroll} at start → \textit{mouseMovement + click + keypress + delete}. \textcolor{brown}{“Refined” editing (specific areas).}                                                            \\
            \midrule
            \multicolumn{2}{c}{\cellcolor[HTML]{EFEFEF}\textbf{(e) Users with high overreliance: Pausing and Hesitation Before LLM Prompting}}                                                                                                                                                      \\
            \multicolumn{2}{c}{\cellcolor[HTML]{EFEFEF}Task: Quiz solving, Window: 50s, $N=9$}                                                                                                                                                                                                      \\
            \cmidrule(lr){1-2}
            \cellcolor[HTML]{FFFFFF}High                  & \cellcolor[HTML]{FFFFFF}Long \textit{idle} on Task page → \textit{keypress + delete} on LLM page. \textcolor{brown}{Initially hesitate, then trust \& adopt \llmres{}.}                                                                 \\
            \bottomrule
        \end{tabular}
    \caption{Summary of User Interaction Behaviors, Interpretations, and Cognitive Processes Associated with Different Levels of \olm{}.}
    \Description{
    This table details five distinct user behavior patterns and contrasts the actions of users with high overreliance versus low overreliance. The table is structured by pattern, each with sections for "High" and "Low" overreliance behavior.
    (a) Frequency of Copying-Pasting: (Task: Trip planning, Window: 10-60s, N=32)
    High: "Repeated copy + paste, no text editing. Adopt LLM's responses directly."
    Low: "Keypress/delete before/after pasting, less copy + paste. Cautious copying + post-paste editing."
    (b) Focused Task Comprehension at the Start: (Task: Article summarization, Window: 10-50s, N=22; Trip planning, Window: 10-50s, N=13)
    Low: "Consecutive alternating mousewheel + scroll at task start. Carefully read task initially."
    (c) Frequency of Referring LLM's responses: (Task: Quiz solving, Window: 10s, N=25)
    High: "Alternated between LLM \& Task pages, single mousewheel + mouseMovement on LLM then click on Task each time. Refer to LLM's responses per ranking."
    Low: "Scroll (mousewheel) on LLM page, then switched to Task page \& do mouseMovement + click series. After referring to LLM's response, finish all the rankings independently."
    (d) Coarse- vs. Fine-Grained Locating \& Editing: (Task: Trip planning, Window: 10-60s, N=36)
    High: "Consecutive mousewheel/scroll + mouseMovement, lengthy copy-paste at end. "Rough" navigation (prominent areas)."
    Low: "Few mousewheel/scroll at start → mouseMovement + click + keypress + delete. "Refined" editing (specific areas)."
    (e) Users with high overreliance: Pausing and Hesitation Before LLM Prompting: (Task: Quiz solving, Window: 50s, N=9)
    High: "Long idle on Task page → keypress + delete on LLM page. Initially hesitate, then trust \& adopt LLM's responses."
    }
    \label{tab:summery_of_quantitative_analysis}
\end{table*}

In this section, we further analyze the 18 salient clusters associated with high or low overreliance, and examine consistently recurring action sequence patterns across the broader set of 54 core valid clusters. For each interaction pattern, we first introduce the \textbf{\textit{Frequency of the Feature}}, followed by \textbf{\textit{Behavioral Characteristics}}---including the action sequence (\actionbox{}) and a behavioral interpretation based on participants’ self-reported decision-making strategies (\behaviorbox{}). Then we make \textbf{\textit{Interpretation}} of the possible underlying cognitive states.

For better illustration, we summarized the results of our quantitative analysis in Table \ref{tab:summery_of_quantitative_analysis}, and visualized the simplified action sequence patterns in Figure \ref{fig:behavior_sequence_all}; raw action sequences are provided in Appendix \ref{Appendix:RawData_from_Behavior_Clustering}. The implementation details of visualization is also included in Appendix. \ref{Appendix:Raw_Data_Pattern_method}.

\begin{figure*}
    \centering
    \includegraphics[width=0.93\linewidth]{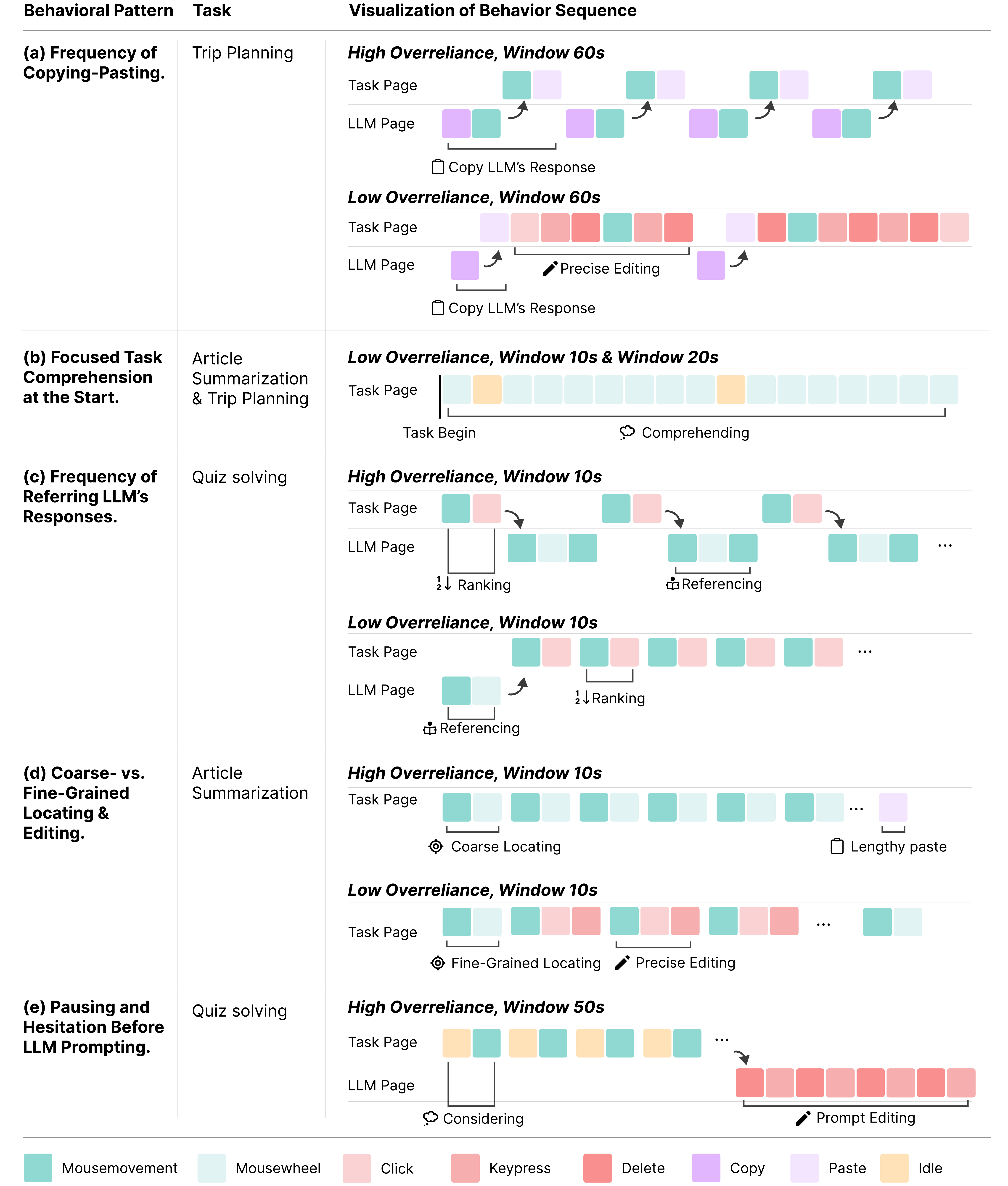}
    \vspace{-1em}
    \caption{Visualization of five simplified action sequence patterns, with three columns respectively denoting Behavioral Pattern, Task, and Visualization of Behavior Sequence. Sub-figures (a) to (e) correspond to distinct behavior patterns: (a) Frequency of Copying-Pasting: In \expThree{}, \hol{} frequently copy/paste unedited, while \lol{} cautiously edit (keypress/delete). (b) Focused Task Comprehension at the Start: In \expTwo{}, \lol{} focus on reading (mousewheel/idle) on the Task page initially. (c) Frequency of Referring \llmres{}: In \expOne{}, \hol{} frequently refer to LLM, while \lol{} refer once then complete tasks independently. (d) Coarse- vs. Fine-Grained Locating \& Editing: In \expThree{}, \hol{} use rough locating (lengthy copy/paste), while \lol{} do precise editing (repeated mouse movement, keypress, click post-scrolling). (e) Pausing and Hesitation Before LLM Prompting: In \expOne{}, \hol{} repeatedly edit (keypress/delete) on LLM page after idling on the Task page.}
    \Description{
    This figure visually represents the five behavior patterns from Table 3. Each pattern shows two timelines of colored blocks, one for "High Overreliance" and one for "Low Overreliance." The blocks represent different user actions, as defined in a key at the bottom.
    Key:
    Mousemovement: Light blue
    Mousewheel: Light green
    Click: Red
    Keypress: Darker pink
    Delete: Gray
    Copy: Purple
    Paste: Light pink
    Idel (Idle): Light orange
    Visualizations:
    (a) Frequency of Copying-Pasting:
    High Overreliance: Shows repeated sequences of purple ("Copy") and light pink ("Paste") blocks on the LLM Page.
    Low Overreliance: Shows a single purple "Copy" block followed by light pink "Paste" and then darker pink "Keypress" blocks, labeled "Precise Editing."
    (b) Focused Task Comprehension at the Start:
    Low Overreliance: Shows a long sequence of light green ("Mousewheel") and light orange ("Idle") blocks on the Task Page at the beginning, labeled "Comprehending."
    (c) Frequency of Referring LLM's Responses:
    High Overreliance: Shows a pattern of alternating between the LLM page and Task page. On the Task Page, there's a single click ("Ranking"), then a switch to the LLM page for a brief action ("Referencing"), and this repeats.
    Low Overreliance: Shows a longer initial action on the LLM page ("Referencing"), followed by a sustained series of actions on the Task Page ("Ranking"), indicating independent work after checking the LLM.
    (d) Coarse- vs. Fine-Grained Locating \& Editing:
    High Overreliance: Shows long sequences of light blue ("Mousemovement") and light green ("Mousewheel") blocks, labeled "Coarse Locating," followed by a large light pink "Lengthy paste" block.
    Low Overreliance: Shows a more varied sequence of actions, including mouse movement, keypress, and delete, labeled "Fine-Grained Locating" and "Precise Editing."
    (e) Pausing and Hesitation Before LLM Prompting:
    High Overreliance: Shows a long light orange ("Idle") block on the Task Page, labeled "Considering," followed by keypress/delete actions on the LLM page, labeled "Prompt Editing."
    }
    \label{fig:behavior_sequence_all}
\end{figure*}

\subsection{Frequency of Copying-Pasting}
\textbf{\textit{Frequency of the Feature.}}
\label{subsubsec:copy_paste_frequency}
This pattern frequently appears in the high overreliance clusters across window 10-60s for \expThree, which covers 36 participants.

\textbf{\textit{Behavior Characteristics.}}
Two distinct behavior sequences were identified (As shown in Figure \ref{fig:behavior_sequence_all}.a). \actionbox{} The copy-paste action consists of a \textit{highlight} and \textit{copy} on the Task/LLM page, a \textit{mouseMovement} to the other page, and then a \textit{paste} there.

\textit{\holup{}}: \behaviorbox{} users usually perform 5–6 copy/paste actions within a 60s window. The copied content is often lengthy, usually consisting of an entire paragraph
\textit{\lolup{}}: \behaviorbox{} users usually perform 1-3 copy/paste actions within a 60s window. The \textit{paste} action is sometimes followed by \textit{keypress} and \textit{delete} action to revise the pasted content.

\textbf{\textit{Interpretation.}}
For \hol{}, the high copy-paste frequency reflects rapid, low-effort uptake of \llmres{}: users may bypass deliberate information screening and reviewing, aligning with System 1’s core attributes-automaticity, minimal deliberation, and intuitive ``quick uptake \cite{kahneman_thinking_2012}.'' In contrast, \lol{}'s behavior align with  System 2, with more careful thinking before adopting \llmres{}.

\subsection{Focused Task Comprehension at the Start}
\label{subsubsec:focused_task_comprehension}
\textbf{\textit{Frequency of the Feature.}}
In terms of timing after task initiation, this behavior occurs at two time points in \expTwo{}: \(17.4 \pm 8.66\) seconds and \(167.41 \pm 18.62\) seconds. In \expThree{}, it is observed around \(16.03 \pm 11.11\) seconds after task initiation.

This pattern was observed across the 10s-50s window in both \expTwo{} and \expThree{}, covering 22 participants. In \expTwo{}, it occurred during the 10s–30s window under both low overreliance and non-significant overreliance conditions; during the 40s–50s window, it appeared only among \lol{}. In \expThree{}, the pattern consistently appeared only in the low overreliance condition across the 10s-50s window, covering 11 participants.
Notably, this pattern never appeared in clusters with high overreliance scores.

\textbf{\textit{{Behavior Characteristics.}}}
As shown in Fig. \ref{fig:behavior_sequence_all}.b, a distinct behavior sequence was identified among \lol{}.

\textit{\lolup{}:}
\actionbox{}
Users performed only \textit{scroll} or \textit{mousewheel} actions on the Task page with occasional \textit{idle} time in between.
\behaviorbox{}
We interpret it as users engaging in focused reading of the task content at the start of the task, using this phase to independently reason about possible answers and form initial expectations before relying on \llmresSingle{}.
For instance, P2048 (a low overreliance user) noted, ``I kinda had an idea in mind of what I wanted to say and I modified GPT's responses to my own. Then I used GPT to figure out word count and to correct any minor grammatical mistakes.''
Similar sentiments emerged in other low overreliance users’ accounts, such as ``I had a rough outline in my mind'' (P1002) and ``I try to imagine if I truly travel to Copenhagen and explore the things I am interested in'' (P1080)---both reflecting a pattern of establishing independent cognitive frameworks (e.g., predefined ideas, preliminary outlines, scenario visualization) prior to engaging with the LLM.

\textbf{\textit{Interpretation.}}
This early, LLM-free engagement with the Task page may reflect a planning or forethought phase: users first analyze the task, activate prior knowledge, and set goals before executing solution strategies, which echo with metacognitive regulation~\cite{panadero2017review}.

\subsection{Frequency of Referring \llmres{}}
\label{subsubsec:pic_interpreting_vs_following}
\textbf{\textit{Frequency of the Feature.}}
This pattern was observed in \expOne{}, where participants ranked items according to their importance via form checkboxes. The relevant clusters across both short (10 seconds) and long (60 seconds) window sizes. Specifically, the high-overreliance pattern appeared in one cluster containing 7 users during the 10th window;
the low-overreliance pattern was observed in window 10, covering 17 participants.

\textbf{\textit{Behavior Characteristics.}}
As shown in Figure \ref{fig:behavior_sequence_all}.c, two distinct behavior sequences were identified:

\textit{\holup{}:} \actionbox{} Users alternated between the LLM and Task pages. For each ranking action, they performed one \textit{mousewheel} action on the LLM page, followed by a single \textit{mouseMovement} and \textit{click} on the Task page. The window switch reached 3–4 times in the 10-second window.
\behaviorbox{} This shows whenever \hol{} encountered a problem during the task, they turned to the LLM for help, and adopted \llmres{} with minimal independent thinking.

\textit{\lolup{}:} \actionbox{} Users scrolled (\textit{mousewheel}) on the LLM page at the beginning of the task, then switched to the Task page and performed a series of \textit{mouseMovement} and \textit{click} actions to complete ranking independently.
\behaviorbox{} This suggests \lol{} first reviewed and comprehended the \llmresSingle{} before acting. They reference \llmres{} only when necessary, and spending more time leveraging their own understanding of the task to complete it.
For example, P2001, a low overreliance participant tended to proactively ask about task-related details to build an understanding of the task before carrying it out independently, noting: ``I asked the AI about what the conditions were like on the moon and how I could utilize each item. I used this information to rank the list of tools.''
In contrast, some high overreliance participants tended to directly ask the LLM for the final answer and then follow it accordingly, mentioning that ``show the situation and the items to GPT, ask she to order them. Then discuss with her about some controversial points. Finally output the list and order,'' ``I inputted the prompt into GPT and read through its answer and reasoning, then asked it to summarize the explanation into a list of the 15 items. I then ranked the items on the task sheet according to GPT's list,'' ``It seemed to provide somewhat accurate answers'', so they ``kept most of gpt's ranking the same'' (P2056, 1015, 2027, 2211, 2148).

\textbf{\textit{Interpretation.}}
Aligned with System 2 processing in dual-process theory, \lol{}'s strategy maintains independent reasoning by first validating and internalizing \llmres{}, then applying it within users’ existing understanding of the task\cite{kahneman2011thinking}. In contrast, \hol{}’s frequent, on-demand consultations with the LLM are consistent with reactive cognitive offloading, where users bypass the effort of independent thinking and instead lean on timely guidance from the AI.

\subsection{Coarse- vs. Fine-Grained Locating \& Editing}
\label{subsubsec:coarse_vs_fine_grained_locating}
\textbf{\textit{Frequency of the Feature.}}
This pattern frequently appears in the high-overreliance clusters across 10–60 s windows for \expTwo{}, which covers 36 participants.

\textbf{\textit{Behavior Characteristics.}}
As shown in Figure \ref{fig:behavior_sequence_all}.d, two distinct behavior sequences were identified:

\textit{\holup{}:}
\actionbox{}
Users performed rapid alternating \textit{mousewheel} (or \textit{scroll}) and \textit{mouseMovement} on the Task page, followed by a lengthy \textit{copy} or \textit{paste} action.
\behaviorbox{}
The rapid interleaved  alternation of events resulted from simultaneous scrolling and cursor positioning.
This pattern shows \hol{} already know the general target (e.g., paragraph start, text input field) and adjust the cursor mid-scroll instead of waiting for scrolling to end. Thus, the target is typically a prominent page area rather than a precise spot (e.g., a specific word). We interpret this as \hol{} adopting ``rough'' editing behaviors, focusing on task completion by navigating to fixed page locations (e.g., copying full task content from the start or pasting a \llmresSingle{} at the end).
Several high-\olm{} participants reported simply inputting the entire article into GPT, giving the summary a quick or ``rough'' check, and then using it as-is because it was ``basically consistent with the article'' or a ``solid summary'' that met the word-count requirement (P1090, P2036, P2052, P2056). For example, P1090 (high overreliance) note: ``I input the article into GPT and ask it to summarize. After a rough review, the summary content is basically consistent with the article.''

\textit{\lolup{}:}
\actionbox{}
Users performed discrete \textit{mouseMovement}, \textit{click}, \textit{keypress}, and \textit{delete} on one page.
\behaviorbox{}
Actions are sequential and not interleaved rapidly, which is interpreted as fine-grained editing.
Specifically, the users scroll to the target area and precisely position the cursor, then conduct ``refined'' edits via repeated mouse movements, keypresses, and clicks to adjust content with precision, prioritizing accuracy over speed.
This represents the fine-grained editing of pasted \llmres{}. For example, P2148 (low overreliance) note:``...I fixed its response by tweaking it based on my own understanding of the article''
Another recurring behavior is to complete an initial version of the task independently and then selectively incorporate small portions of \llmresSingle{} into that draft. For instance, P2140, P2115 note that they ``mostly'' writing the summary themselves while ``used some of GPT’s sentences as well''.

\textbf{\textit{Interpretation.}}
These locating \& editing patterns indicate different ways that users regulate their cognitive effort when interacting with \ai{}. The coarse, chunk-level navigation resemble a form of cognitive offloading, where \hol{} minimize internal effort by transporting large blocks of \llmres{} into the Task page with little subsequent modification~\cite{gilbert2020optimal}.
In contrast, we interpret the fine-grained locating and micro-edits are consistent with effortful, analytic adjustment of external representations: users first construct or maintain their own mental model of the answer, then use \llmres{} as material to be selectively inserted and corrected at the level of specific sentences or phrases, aligning more closely with System~2-like processing and metacognitive control~\cite{evans2013dual,nelson1990metamemory,metacognitive2024tankelevitch}.

\subsection{Pausing and Hesitation Before LLM Prompting}
\label{subsubsec:deliberating_vs_wavering}
\textbf{\textit{Frequency of the Feature.}}
This pattern is observed in the high overreliance clusters of window 50 in \expOne{}, covering 9 participants. It never appeared in the clusters with low overreliance scores.

\textbf{\textit{Behavior Characteristics.}}
As shown in Fig. \ref{fig:behavior_sequence_all}.e, a distinct behavior sequence was identified among \hol{}.

\textit{\holup{}:}
\actionbox{}
This pattern is characterized by a long \textit{idle} on the Task page, followed by \textit{keypress} and \textit{delete} actions on the LLM page.
\behaviorbox{}
We interpret this pre-prompt-editing pause as a sign that users are cognitively engaged with the \llmres{}, reflecting processes like questioning or reconsidering their own input and the \llmres{}. When they suspect inaccuracies in the \llmres{}, they nonetheless continue interacting with the \ai, using \textit{keypress} and \textit{delete} to craft follow-up questions. Overtime, users may prone to be persuaded by \ai{}.
For example, P1014 notes, ``First, I asked GPT to provide the results directly. Then, I discussed with it the reasons for my disagreement, and it convinced me.''
Several high-overreliance participants also described strategies that involved actively “discussing” controversial points with the LLM or “asking GPT any questions I had”, or ``asked follow-up questions regrading the differences in our answers'' (P1043, P2152,P1006). This follow-up behavior is validated by our analysis of input within the cluster, and examples include questions like ``Why are salt tablets useful?'' or ``But there is no air on the moon?''.

\textbf{\textit{Interpretation.}}
We interpret this behavior as stemming from metacognitive monitoring without control ~\cite{metacognitive2024tankelevitch,vasconcelos2023explanationsreduceoverrelianceai,10.1145/3449287}~\footnote{Classical models distinguish metacognitive monitoring (assessing the state of one's cognition) from metacognitive control (the ability to use those judgments to alter behavior)~\cite{son2002relation}.}.
Prior work suggests that monitoring and control can dissociate and people may experience a ``metacognitive friction''~\cite{gilbert2020optimal,crystal2011evaluating}. This is consistent with our experiment, where participants might sense ``something might be wrong in \llmres{}'' (metacognition monitoring), yet failed to seek external sources for verifications (metacognitive control).
Potential reasons include time pressure \cite{swaroop2024accuracy}, weak motivation, the persuasive influence of AI-generated suggestions~\cite{dodge2019explain}, or cost-benefit tradeoffs ~\cite{nelson1990metamemory, gilbert2020optimal}.

\section{Discussions}

Existing methods for detecting overreliance primarily evaluate task outcomes~\cite{Kim2024, bo2025relyrelyevaluatinginterventions,dey2025knowmistakespreventingoverreliance}, which is appropriate for traditional AI classification tasks. For \ai, however, such \textit{outcome-oriented indicators} are limited: users’ goals may be implicit, intermediate turns may not map cleanly to final correctness, and errors made during interaction are often difficult to attribute to specific causes. Moreover, relying on a hallucination-prone LLM to judge the correctness of a potentially erroneous final answer creates a logical paradox.

Motivated by the recurring nature of overreliance-related behaviors across tasks, our contribution is to shift from \textit{outcome-based evaluation} to \textit{process-level behavioral characterization}. We identify five interaction patterns correlated with different levels of \olm{} across three representative \ai{} tasks. The observable behavior patterns and the autoencoder–clustering framework in Section~\ref{subsubsec:post_clustering_process} can also be adapted for real-time detection and intervention.

The remainder of this section discusses factors that may moderate the relationship between behaviors and \olm{} (Section~\ref{subsec:factors_moderators}); design implications for mitigation strategies (Section~\ref{subsec:design_recommendations}); and limitations and future directions, including misinformation simulation validity, generalizability, data expansion, and cognitive validation (Section~\ref{subsec:limitation_futurework}).

\subsection{Factors that Influence Behaviors}
\label{subsec:factors_moderators}

While our analysis identifies several behavioral patterns associated with overreliance, these behaviors may also arise from alternative factors. We discuss three major classes of confounding factors, task-, LLM-, individual-related factors, that may also shape user behaviors in ways that obscure overreliance.

\subsubsection{Task-related factors}
Task demands such as difficulty and time pressure can substantially influence how users allocate attention and effort. Prior work shows that high task difficulty or tight deadlines push users toward shallow processing and rapid decision strategies, even in the absence of AI support~\cite{evans2005rapid, brisson2014belief, neys2006dual}. In our study, the 15-minute limit and the complexity of the tasks may have induced such efficiency-oriented behaviors. Behaviors such as rapid acceptance of plausible answers or minimal verification could therefore also reflect time-management strategies.

\subsubsection{LLM-related factors}
Interacting with LLMs adds prompt formulation, output parsing, and context maintenance, which increases cognitive load~\cite{metacognitive2024tankelevitch} and can lead users to simplify their workflow regardless of how much they trust the model. Second, LLMs encourage cognitive offloading similar to search engines or navigation tools~\cite{sparrow2011google, RISKO2016676}.

In addition, the fluent, authoritative, single-answer style of \llmres{} can reduce scrutiny and anchor users on the first answer they see~\cite{spatola2024efficiency}, producing behaviors that resemble \olm{} even among cautious users.
\subsubsection{User-related factors}
Individual differences can significantly shape behavioral traces. Users vary in metacognitive knowledge about LLM limitations~\cite{metacognitive2024tankelevitch}; some may not recognize that LLMs can hallucinate, leading them to copy answers without verification, which reflects limited awareness. Users also vary in motivation: low-motivation participants may appear overreliant simply because they minimally engage with the task (cognitive miserliness), while highly motivated users may override or scrutinize LLM output even when unnecessary. Expertise similarly modulates behaviors. Novices face greater information asymmetry and are more easily persuaded by fluent or technical-sounding responses~\cite{metacognitive2024tankelevitch}, making them more likely to accept incorrect outputs, not necessarily because of misplaced trust, but because they lack the basis for evaluation.

These factors highlight the need for future work to experimentally vary these factors to separate genuinely overreliant behaviors from rational strategies and interface- or user-driven effects.

\subsection{Design Recommendations for Conversational LLMs' Interfaces}
\label{subsec:design_recommendations}

Our findings suggest five behavioral patterns that correlate with higher \olm{}. We propose design recommendations that treat each pattern as a soft behavioral trigger for offering just-in-time mitigation.

\paragraph{Design for ``High-Frequency Copy-Pasting'': Context-Aware Auto-Verification.}
Trigger: Users repeatedly copy and paste \llmres{} into the working document.
Mitigation: The system can automatically highlight or tag key factual elements being pasted (e.g., named entities and numbers) to nudge user verification.
To enhance explainability, the backend can further retrieve corroborating evidence from external sources, compute a confidence score for each highlighted element with relevant sources, referring to the approaches such as HILL~\cite{leiser2024hill}.

\paragraph{Design for ``Insufficient Task Comprehension'': Adaptive Task Road-map + Task Confirmation}

Trigger: Users spend very little time on the task page before issuing their first prompts.
Mitigation: The system proactively produce a floating panel including a list or a flowchart to guide users to understand the task goals.
The system could also present a ``Task Confirmation Window'' that ask users to check or edit a short restatement of goals and constraints before prompting.

\paragraph{Design for ``High-Frequency of Referring \llmres{}'': Task Chunking}

Trigger: Within a short time window, users repeatedly switch between the task document and the LLM interface.
Mitigation: The system can invite users to list or draft all questions they anticipate asking \ai{}. Then the system reorganizes these questions into a structured task-questions outline, helping users manage them in a modular way, such as ``execute all at once'' or ``start with high-level questions first.''

This design echoes chunking strategy \cite{nassar2018chunking, thalmann2019does} by compressing fragmented queries into manageable units, thus reducing users’ cognitive load needed to remember all the queries.

\paragraph{Design for ``Coarse-Grained Locating \& Editing'': Granular Review Staging+Interaction Frictions}

Trigger: Users navigate with rapid, coarse scrolling and editing.
Mitigation:
The system can temporarily segment the pasted text into smaller, granular modules and offer quick checks before users apply the full block.
The granularity of these modules is adjusted based on users’ detected locating \& editing pattern, with more coarse-grained behaviors triggering finer module segmentation (e.g., sentence-level module versus paragraph-level module).
Furthermore, when more frequent coarse-grained locating and editing are detected, the system can restrict scroll ranges to a maximum of one quarter to one half per second.

\paragraph{Design for ``Pausing/Hesitation Before LLM Prompting'': Query Visualization+Cognitive Forcing Functions}

Trigger: Users input repeated, highly similar prompts to \llmres{}.
Mitigation: the system can visualize the recent query trail to surface repeated queries and potential ``stuck'' states, suggest alternative resources (e.g., search, domain references, manual notes), or momentarily switch to a more Socratic style of response that asks clarifying questions instead of providing direct answers.

Caveat: These design recommendations should be complemented by contextual factors (refer to Section~\ref{subsec:factors_moderators}), which allow calibrating detection thresholds and risk levels, since identical behaviors may imply different risks for different users or scenarios.

\subsection{Limitations \& Future Work}
\label{subsec:limitation_futurework}

\subsubsection{Simulating LLM-Generated Misinformation}
\label{subsec:discussion_time}

We acknowledge several limitations related to ecological validity and generalizability of our misinformation design:

\textbf{Natural hallucination vs. prompt injecting:} Regarding ecological validity, we acknowledge that artificially injected misinformation does not fully capture the stochastic nuances of naturally occurring hallucinations. Without established method to fully mimicking real-world hallucinations, we followed prior work~\cite{kim2024m, zhou2023synthetic} by using prompting as a pragmatic approach and carefully maintain fluency, style, and contextual consistency (Section~\ref{subsec:design}) to ensure the misinformation appears natural rather than blatantly false, which was confirmed in pilot tests.

\textbf{Prompting vs. fine-tuning:} While fine-tuning may produce subtler or more contextually consistent errors, prior evidence suggests that prompting and fine-tuning yield comparable results on many general tasks~\cite{yin2024deeper, yang2025adversarial}, and no comprehensive study currently compares them for misinformation generation with human evaluation.

\textbf{Consistent responses vs. model self-correction:} Our design choice that the LLM would consistently provide a single (misleading) statement when challenged has a degree of ecological validity. While modern LLMs can self-correct, prior work~\cite{laban2025llms} shows that even state-of-the-art models often stick to initial responses despite errors (e.g., the ``9.11 is larger than 9.9'' example). Given the prevalence of AI- or human-generated misinformation, users persisting with and adapting around a stable but incorrect response reflects realistic interaction dynamics beyond our specific prompting setup.

\subsubsection{Generalizability Across Tasks}

Our study is primarily limited by the specific nature of the experimental tasks: focusing on overreliance driven by misinformation in a constrained editing scenario. Consequently, the generalizability of our findings to broader contexts remains to be verified. Future work could expand to a wider range of tasks across three key dimensions:

First, regarding task complexity and openness, future research should move to open-ended and generative tasks (e.g., creative writing, complex reasoning, or code generation). In these scenarios, overreliance may not manifest simply as factual errors, but as more subtle issues such as diminished creativity, homogenized content, or logical inconsistencies, which are difficult to capture through simple outcome metrics.

Second, regarding task stakes, our study involves low-risk task; however, overreliance is particularly critical in high-stakes decision-making tasks (e.g., medical diagnosis, legal analysis, or financial forecasting). In such domains, users might defer to LLM not to save effort, but to shield themselves from the accountability of potential errors.

Finally, this expansion of task diversity exposes the limitations of current evaluation methods. Since our study assesses overreliance based on final performance outcomes (e.g., error rates), it cannot capture the overreliance changes during the process. Future work could develop real-time detection methods that monitor users’ behavior throughout the task. For example, by measuring scrutiny time, acceptance latency, or incorporating physiological indicators to infer users’ moment-to-moment overreliance levels. Such methods would enable a more nuanced understanding of how overreliance evolves in response to factors such as time, task content, and external stimuli.

\subsubsection{Lack of Cognitive Truth}

We interpret participants' cognition by mapping each behavioral pattern to the cognitive theories whose core features closely align with the behavioral characteristics, and we reinforce these interpretations by cross-validating them with participants’ strategy reports.
We acknowledge that current methods have limitations. Strategy reports are collected after task completion and often reflect overall impressions rather than behavior-specific reasoning. Moreover, since users self-report their strategies, descriptions are sometimes vague or incomplete, focusing on their attitudes or feelings toward the LLM rather than detailed behavioral rationales.

To avoid contaminating overreliance behaviors, we opted not to rely on concurrent (think-aloud, confidence ratings) or retrospective validation, which prior work shows can induce observer effects~\cite{recallsystematic, hamilton2016audience}. In our pilot tests, think-aloud protocols led participants to either stop talking under unsupervised conditions or increase cognitive load under supervision, while post-task labeling often elicited mere action repetition rather than reasoning explanations.

Future work could incorporate real-time cognitive assessment methods, such as sensing technologies (e.g., eye tracking, electrodermal activity, heart rate, EEG) or controlled experimental manipulations (e.g., dual-task paradigms, time constraints), to more precisely measure cognitive states.

\subsubsection{Data Insufficiency}
Although our work has identified five distinct behaviors that could indicate overreliance, the volume of data we have collected is insufficient to develop a predictive model. To establish a more comprehensive set of behavioral indicators for \olm{}, future research can build on the five behavioral patterns we identified to further mine additional indicator data. For example, to gain deeper insights into how users engage in initial task comprehension, we can record the proportion of text covered by the user’s scroll view on the task page; this metric will help us better understand the extent to which users read and process task instructions. In future work, we plan to collect more data to develop a robust behavioral detection model for \olm{}.

\section{Conclusion}

In this study, we sought to understand overreliance on \ai based on users' interaction behaviors. Through a data collection study and cluster-based analyses, we identified significant correlations between overreliance and specific patterns of interaction behaviors, such as increased copy-paste actions and varying interaction frequencies on Task and LLM pages.

Our work contributes new knowledge to research on \olm{} by providing a dataset of interaction behaviors and establishing a foundation for future research aimed at developing adaptive strategies to detect and mitigate overreliance in real-time. By leveraging users' interaction behaviors, we can implement timely interventions that foster a healthier balance in human-AI collaboration, ultimately enhancing the effectiveness of AI as a supportive tool in various applications.

\begin{acks}
We thank Professor Yingqing Xu and Professor Chun Yu from Tsinghua University for providing guidance and access to necessary research facilities for this work.
\end{acks}

\bibliographystyle{ACM-Reference-Format}
\bibliography{ref,software}

\newpage
\appendix
\section{Appendix}
\subsection{Task Details}
\label{appendix:task_details}
\subsubsection{\expOneCaps~(\taskDurationWithUnit{} for each trials)}

This task operationalizes a real scenario: users tend to rely on AI's misinformation even when their own knowledge would serve them better.

Adapted from NASA survival exercises \cite{larson2019team, nowak2023hear, esfandiari2024meet}, the quizzes ask participants to imagine themselves in a harsh environment and rank 15 items by their importance for survival. Since the task relies on common sense as the original exercise prohibits external tools, and participants complete it without accessing external information.

To mitigate learning effects, we selected two distinct tasks that are comparable in difficulty (survival in desert and on the moon). To control for order effects, we counterbalanced: (1) with and without AI assistance, and (2) the environment type during the 2 trials.

\subsubsection{\expTwoCaps~(\taskDurationWithUnit)}

This task operationalizes a real scenario where users use AI to process information from a specific source (e.g., summarizing, extracting details, or answering questions based on it)---and the AI may generate inaccurate responses due to hallucination.

In this task, participants are required to read a science article and summarize it. We selected an article about sneezing for two key reasons: first, the topic is familiar and interesting to most people, which helps boost their motivation to engage with the task; second, approximately 3\% of its total words include complex vocabulary and concepts, requiring participants to carefully identify misinformation. Since the task only relies on the content of the aforementioned article, participants will be assisted by LLM but are not allowed to use search engines. 

\subsubsection{\expThreeCaps~(\taskDurationWithUnit)}

This task operationalizes a real scenario: AI may provide outdated content---yet over-reliant users may fail to switch tools freely to verify, leading to adoption of outdated content.

In this task, participants are asked to prepare for a trip to Copenhagen by listing specific destinations and their relevant details. As these details are time-sensitive, participants should search the internet for correct information.

\subsection{Adjusting the \expOneCaps}
\label{subsec:adjusting_task_one}
Before implanting misinformation, we conducted a pilot study with 54 participants, which demonstrated significant differences in the final scores between two existing NASA survival exercises \footnote{The method for computing the scores are listed in Section \ref{subsubsec:metric_exp_one}}. 
The score for the desert survival task was significantly higher than that for the moon survival task (p < 0.01), indicating participants perform worse in desert survival task. 

To make the two trials' difficulty comparable, we excluded the three items (food concentrate, two .45 calibre pistols, self-inflating life raft, marked with * in Table \ref{tab:obj_moon}) with the highest average absolute difference $|\texttt{diff}_i|$ from the desert survival score calculation, and removed the three items (bottle of 1000 salt tablets, overcoat for everyone, cosmetic mirror, marked with * in Table \ref{tab:obj_desert}) with the lowest average absolute difference $|\texttt{diff}_i|$ from the moon survival score calculation. After this adjustment, we have $N=12$ items, and the scores between desert survival and moon survival was no longer significant (p=0.08).

To achieve the ranking adjustment, we fixed the order of items that showed minimal ranking differences in the pilot study (to prevent users from quickly detecting that the LLM was providing inaccurate information and consequently disengaging from it). For the remaining items, we adjusted their rankings based on the magnitude of the differences, moving them further from their original positions. The adjusted results are shown in Table \ref{tab:obj_moon} and Table \ref{tab:obj_desert}. Finally, we ensured that when the adjusted final rankings were applied to the choices made by participants in the pilot study, there would be no significant differences in the scores computed in the two different survival exercises.

\begin{table}[]
    \centering
    \begin{tabular}{p{10em}|cccp{3em}}
        \hline
        objects & $index^{gt}$ & $index^{p}$ & $|diff|$ & manual index\\
        \hline
        (1) box of matches & 15 & 11.18 & 3.82 & 3\\
        * (2) food concentrate & 4 & 4.7 & 0.7 & 4\\
        (3) 50 feet of nylon rope & 6 & 8.78 & 2.78 & 14\\
        (4) parachute silk & 8 & 9.34 & 1.34 & 7\\
        (5) portable heating unit & 13 & 8.94 & 4.06 & 1\\
        * (6) two .45 calibre pistols & 11 & 11.98 & 0.98 & 12\\
        (7) one case of dehydrated milk & 12 & 9.54 & 2.46 & 15\\
        (8) two 100 lb. tanks of oxygen & 1 & 3.16 & 2.16 & 8\\
        (9) stellar map & 3 & 6.92 & 3.92 & 11\\
        * (10) self-inflating lift raft & 9 & 9.34 & 0.34 & 9\\
        (11) magnetic compass & 14 & 8.1 & 5.9 & 10\\
        (12) 20 litres of water & 2 & 5.14 & 3.14 & 13\\
        (13) signal flares & 10 & 7.3 & 2.7 & 5\\
        (14) first aid kit, including injection needle & 7 & 8.38 & 1.38 & 6\\
        (15) solar-powered FM receiver transmitter & 5 & 7 & 2 & 2\\
        \hline
    \end{tabular}
    \caption{Objects and Ranking Indices of Moon Survival in Pilot Study}
    \label{tab:obj_moon}
\end{table}

\begin{table}[]
    \centering
    \begin{tabular}{p{10em}|cccp{3em}}
        \hline
        objects & $index^{gt}$ & $index^{p}$ & $|diff|$ & manual index\\
        \hline
        (1) torch with 4 battery cells& 4& 7.64& 3.64&11\\
        (2) folding knife & 6& 6.42& 0.42&7\\
        (3) air map of the area & 12& 5.22& 6.78&6\\
        (4) plastic raincoat (large size)& 7& 9.06& 2.06&14\\
        (5) magnetic compass& 11& 5.1& 5.9&3\\
        (6) first-aid kit& 10& 5.92& 4.08&2\\
        (7) 45 calibre pistol (loaded)& 8& 9.08& 1.08&12\\
        (8) parachute (red \& white)& 5& 9.52& 4.52&15\\
        * (9) bottle of 100 salt tablets& 15& 7.46& 7.54&8\\
        (10) 2 litres of water per person& 3& 3.24& 0.24&4\\
        (11) a book entitled ``Desert Animals''& 13& 9.5& 3.5&5\\
        (12) sunglasses (for everyone)& 9& 8.64& 0.36&9\\
        (13) 2 litres of 180 proof liquor& 14& 10.9& 3.1&1\\
        * (14) overcoat (for everyone)& 2& 9.4& 7.4&10\\
        * (15) a cosmetic mirror& 1& 12.9& 11.9&13\\
        \hline
    \end{tabular}
    \caption{Objects and Ranking Indices of Desert Survival in Pilot Study}
    \label{tab:obj_desert}
\end{table}

\subsection{Details on Computing Scores for \expOneCaps{}}
\label{subsubsec:metric_exp_one}
We begin by introducing the metric calculation method used in the original NASA tests \cite{larson2019team, nowak2023hear, esfandiari2024meet}.
Both trials in \expOneCaps{} require participants to rank the importance of $N=15$ items. Each trial has its standard answer provided by experts. The score for each trial is calculated as follows: first, we compute the absolute difference $|\texttt{diff}_i|$ of item $i$ between its index in the ``groundtruth'' ranking ($\texttt{index}_{i}^{gt}$) and its index in the participant's answer ($\texttt{index}_{i}^{p}$). Then, we sum up all absolute difference values to get the score for one trial (\(score_{trial}\), Equation \ref{equ:task_1_score_1}).

This score reflects the difference between the item rankings provided by the participants and the standard rankings given by experts. A smaller difference indicates better participant performance, while a larger difference suggests poorer performance. 
\begin{equation}
    \text{score}_\text{trial} = \frac{1}{N}\sum_{i=1}^{N}{|\texttt{diff}_i|} = \frac{1}{N}\sum_{i=1}^{N}{|\texttt{index}_{i}^{gt} - \texttt{index}_{i}^{p}|}
    \label{equ:task_1_score_1}
\end{equation}

\subsection{Informed Consent Statement}
\label{subsubsec:informedConsentStatement}
The purpose of this study is to understand interaction behaviors during engagement with LLM. Throughout the experiment, we will collect your mouse and keyboard input data from the browser, as well as your communication records with the LLM. All data will be anonymous, shared only within the project team, and used solely for academic research and publication purposes.

\subsection{Preprocessing Details}
\label{appendix:preprocessing}
Our data logging script recorded events at very short intervals, 

we merged the raw data to improve clarity. Specifically:
\begin{enumerate}
    \item consecutive mouseMovement events were merged into a single event, preserving the total movement distance and the start/end times.
    \item consecutive mousewheel and scroll events were merged based on movement direction, ensuring mousewheel and scroll in the same direction were combined into a single event. We also preserve the distance, direction, and the start/end times.
    \item consecutive keypress events were merged into a single event, concatenating the input characters into a string, and preserving the start/end times.
    \item consecutive delete events were also merged, preserving the number of deletions and the start/end times.
\end{enumerate}

\subsection{Distributions of Participants' Scores in Tasks}
\begin{figure}
    \centering
    \includegraphics[width=1\linewidth]{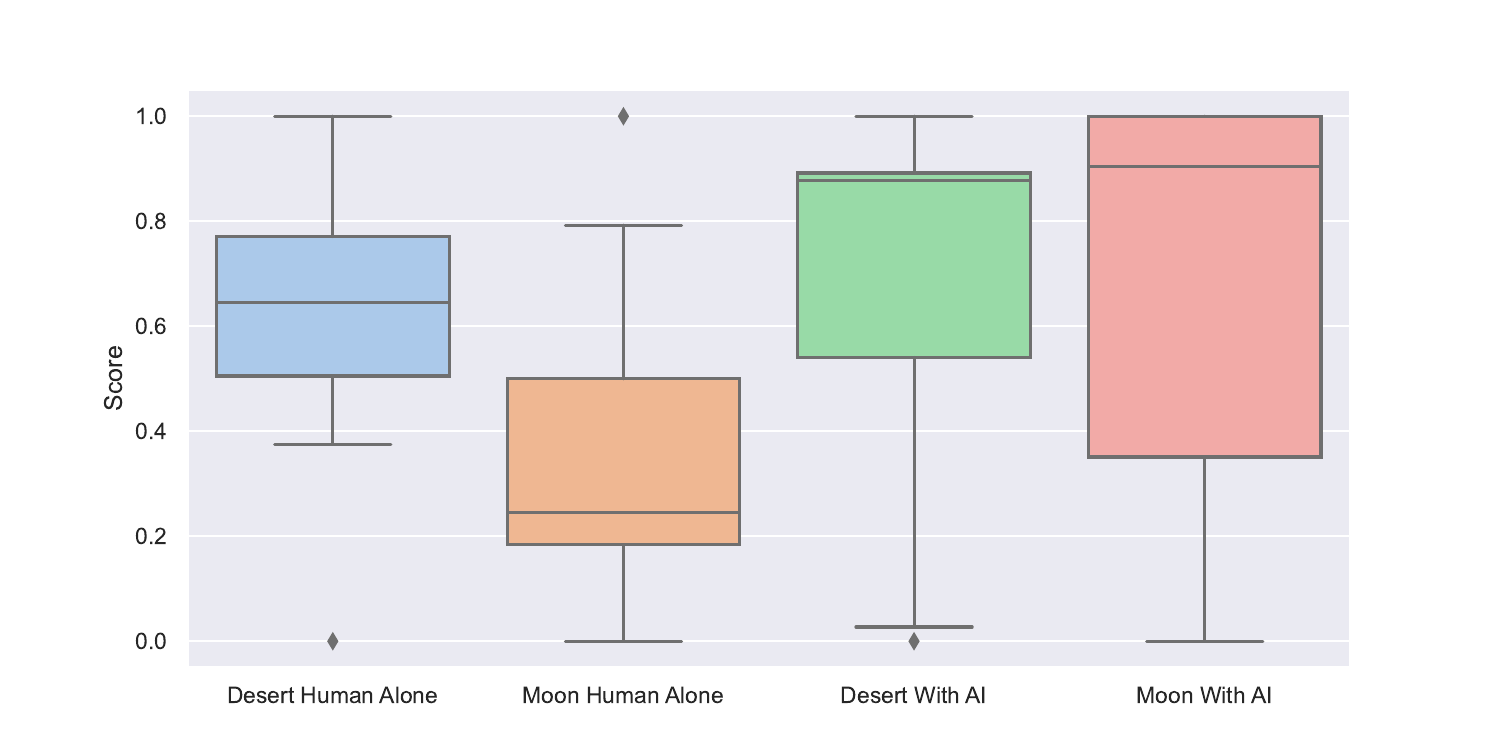}
    \caption{Box plot comparing scores of \expOne{} across four conditions: human alone in the desert, human alone on the moon, with AI in the desert, and with AI on the moon.}
    \label{fig:task_1_score}
\end{figure}

We present the histogram of normalized scores in Figure \ref{fig:histogram_of_normalized_score}. Additionally, the box plot of scores for Task 1 is shown in Figure \ref{fig:task_1_score}.
The results indicate no significant relationship between the scores of desert \withoutAI{} and moon \withoutAI{} $(p>0.05)$. 
Furthermore, the metrics of desert \withAI{} and moon \withAI{} are significantly higher than those of desert \withoutAI{} and moon \withoutAI{} respectively ($p<0.05$).
The significant relationships observed in the results are consistent with our experimental design hypotheses: \one participants would perform similarly in two subtasks, \two participants would perform worse with \ai that generates misinformation. 

\subsection{The Filtering Metrics Values for Different Tasks and Windows}
\begin{figure*}
    \centering
    \includegraphics[width=0.9\linewidth]{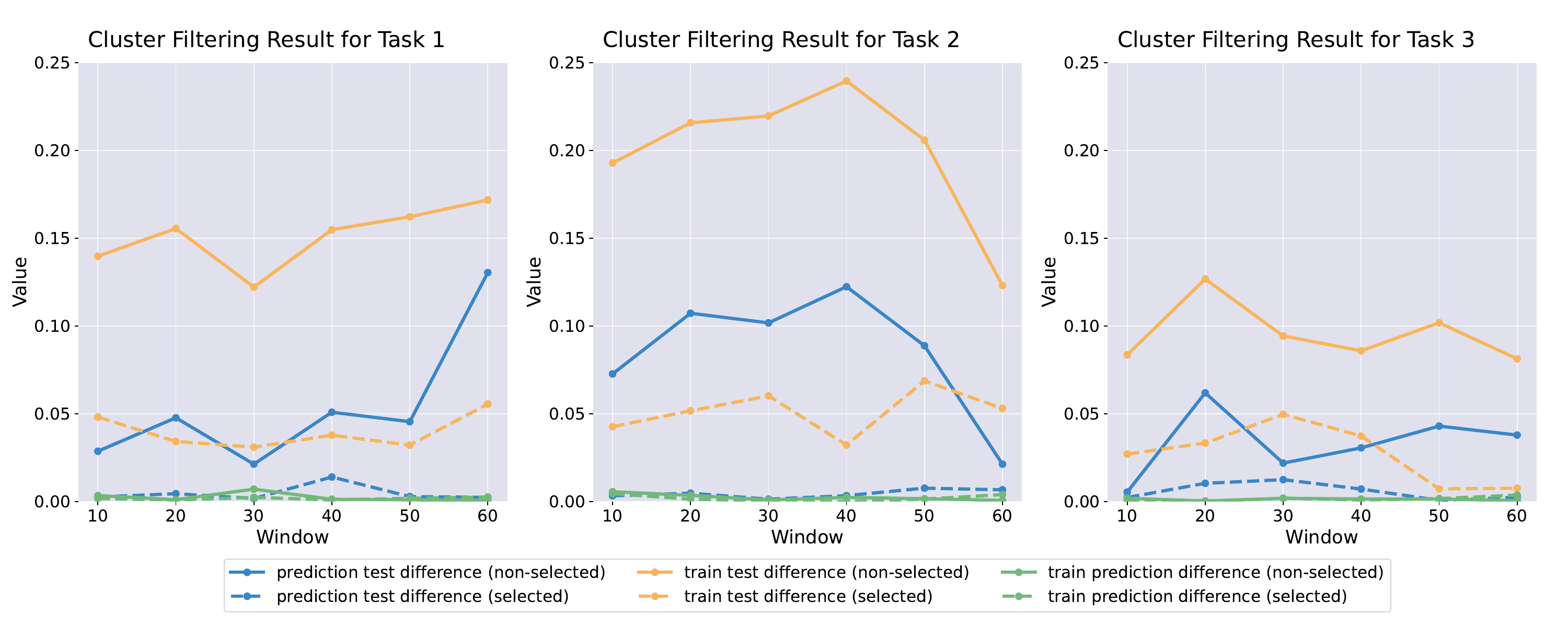}
    \caption{The Filtering Metrics Values for Different Tasks and Windows.
    }
    \label{fig:cluster_filtering_result}
\end{figure*}

We present the filtering metrics values for different tasks and windows in Figure \ref{fig:cluster_filtering_result}.

\subsection{Raw Data from Behavior Clustering}

\subsubsection{Visualization Method for Raw Data}
\label{Appendix:Raw_Data_Pattern_method}
To support this analysis, we first visualized the rough data from behavior clustering.

In each visualization, colored blocks represent different types of actions, arranged vertically to reflect the temporal order from top to bottom. For each action, the part of the label after the underline (``\_’’) indicates the page where the action occurred (either the LLM page or the Task page). Correspondingly, text aligned to the left denotes actions on the Task page, while right-aligned text indicates actions on the LLM page. 
We applied the same visualization format to all the typical sequences (as mentioned in Section 

\ref{sec:quantitative_findings}
) from each selected cluster across different tasks and time windows, organizing them hierarchically by cluster, window, and task to facilitate the subsequent analysis. Although the figure only shows sequences of action types, we will refer to the preprocessed behavioral logs described in Section \ref{subsec:preprocessing} for information such as timestamp or action features when necessary.

\begin{figure}[htbp]
    \centering
    \begin{subfigure}[b]{0.2\textwidth}
        \centering
        \includegraphics[width=\textwidth]{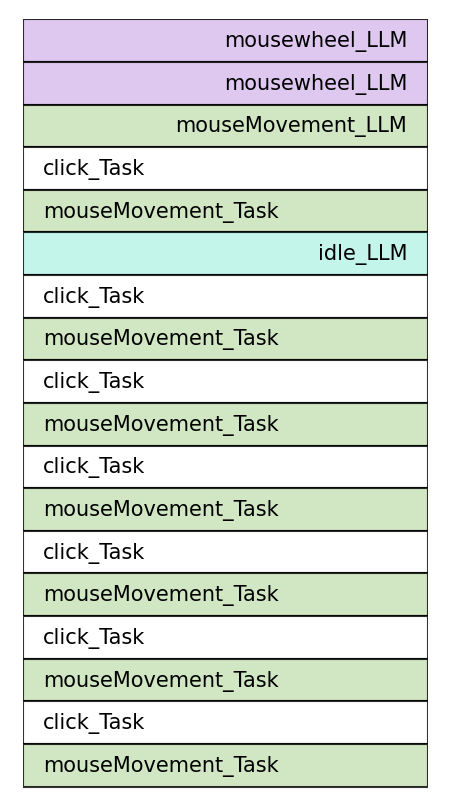}
        \caption{\expOne{}, Window 10, Low Overreliance}
        \label{fig:task1_interpreting_following_low}
    \end{subfigure}
    \begin{subfigure}[b]{0.2\textwidth}
        \centering
        \includegraphics[width=\textwidth]{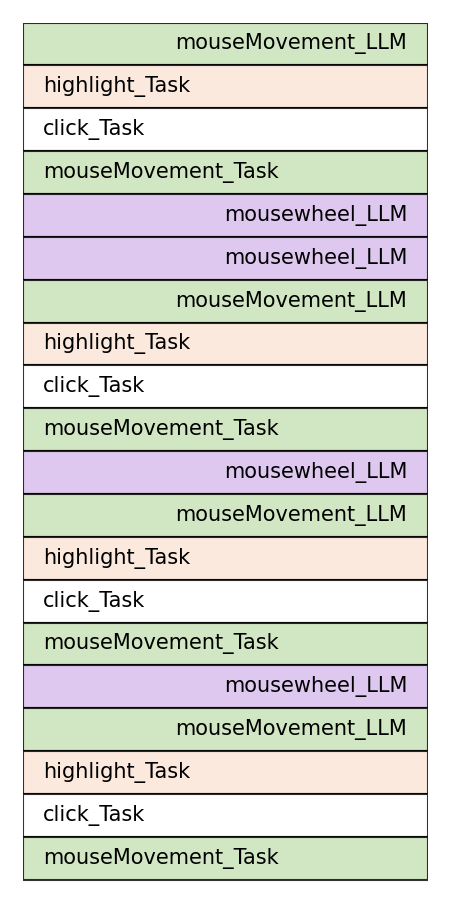}
        \caption{\expOne{}, Window 10, High Overreliance}
        \label{fig:task1_interpreting_following_high}
    \end{subfigure}
    \caption{Two Patterns in \expOne{}, Windows 10 for Interpreting and Following \llmres{}.}
    \label{fig:task1_interpreting_following}
\end{figure}

\begin{figure}[htbp]
    \centering
    \begin{subfigure}[b]{0.2\textwidth}
        \centering
        \includegraphics[width=\textwidth]{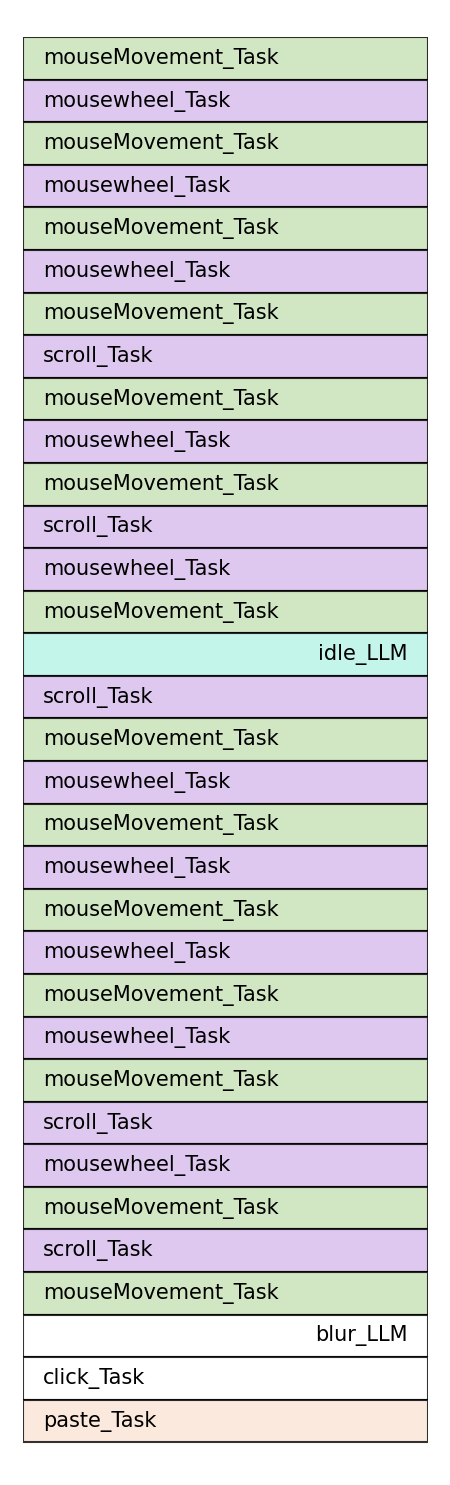}
        \caption{\expTwo{}, Window 50, High Overreliance (Partial)}
        \label{fig:task2_locating_window_50}
    \end{subfigure}
    \begin{subfigure}[b]{0.2\textwidth}
        \centering
        \includegraphics[width=\textwidth]{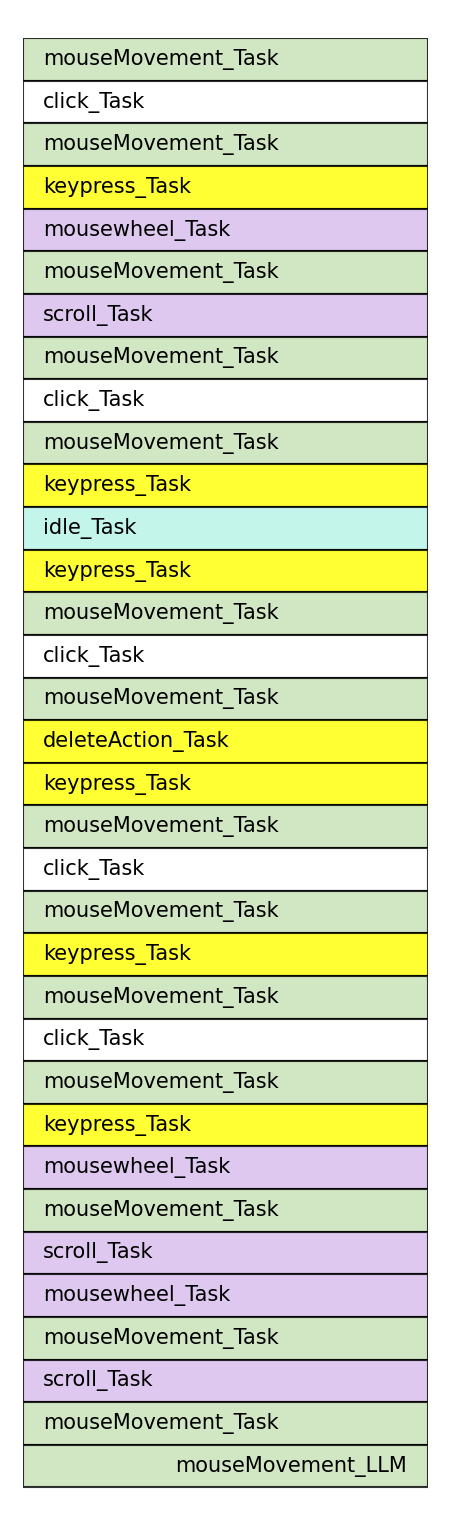}
        \caption{\expTwo{}, Window 60, Low Overreliance (Partial)}
        \label{fig:task2_locating_window_60}
    \end{subfigure}
    \caption{Target Selection for \expTwo{}}
    \label{fig:task2_locating}
\end{figure}

\begin{figure}[htbp]
    \centering
    \begin{subfigure}[b]{0.2\textwidth}
        \centering
        \includegraphics[width=\textwidth]{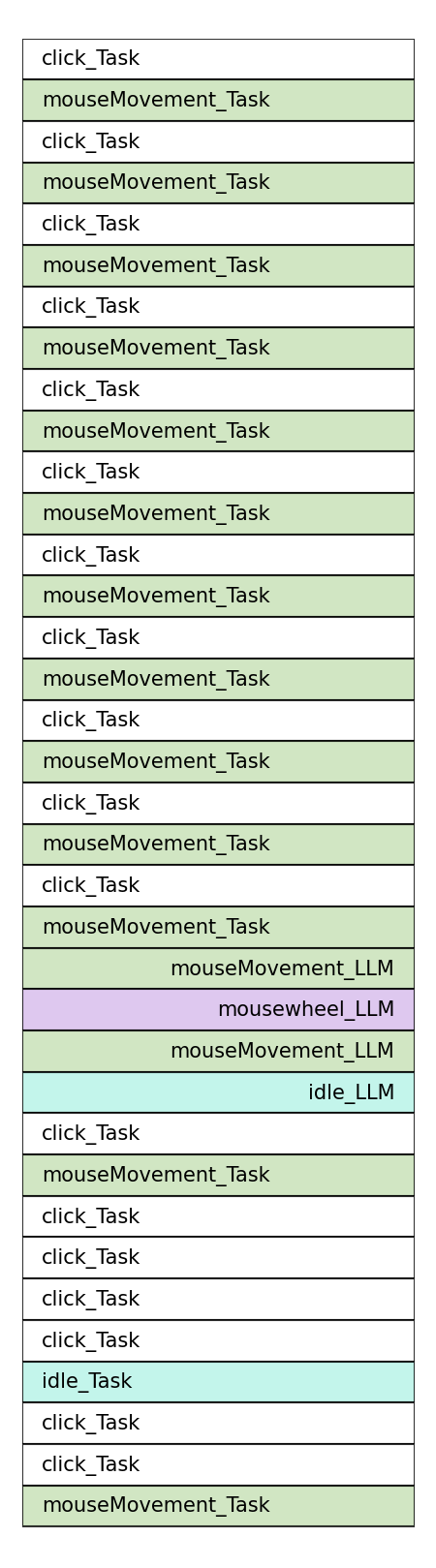}
        \caption{\expOne{}, Window 60, Low Overreliance (Partial)}
        \label{fig:task1_interpreting_following_supplement_window60_low}
    \end{subfigure}
    \begin{subfigure}[b]{0.2\textwidth}
        \centering
        \includegraphics[width=\textwidth]{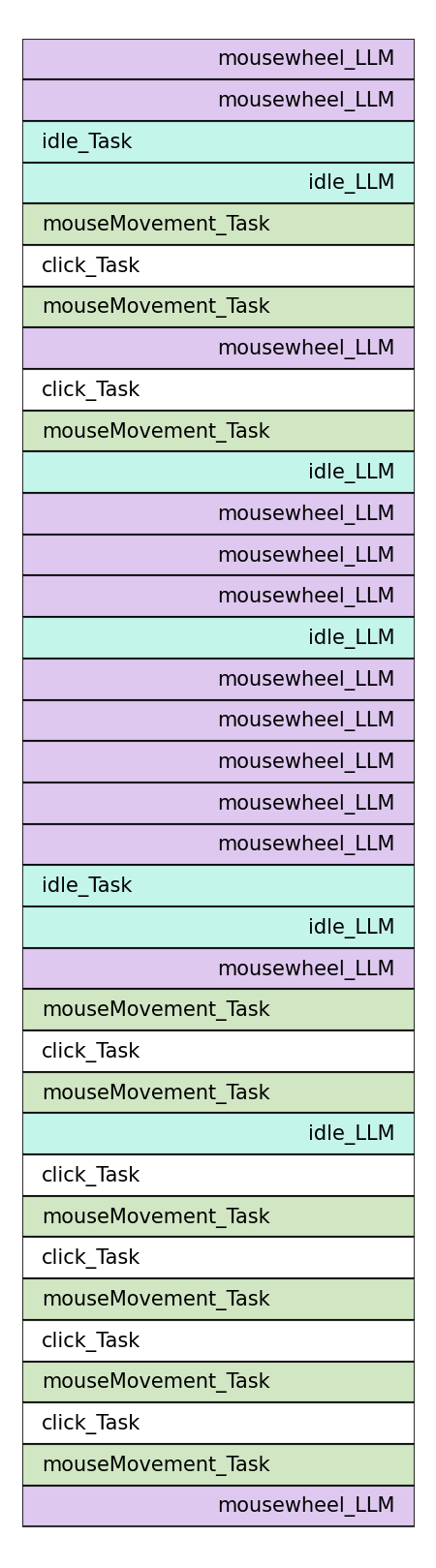}
        \caption{\expOne{}, Window 30, High Overreliance}
        \label{fig:task1_interpreting_following_supplement_window30_high}
    \end{subfigure}
    \caption{Two Patterns in \expOne{} (Supplement) for Interpreting and Following \llmres{}. 
    }
    \label{fig:task1_interpreting_following_supplement}
\end{figure}

\begin{figure}[htbp]
    \centering
    \begin{subfigure}[b]{0.2\textwidth}
        \centering
        \includegraphics[width=0.5\textwidth]{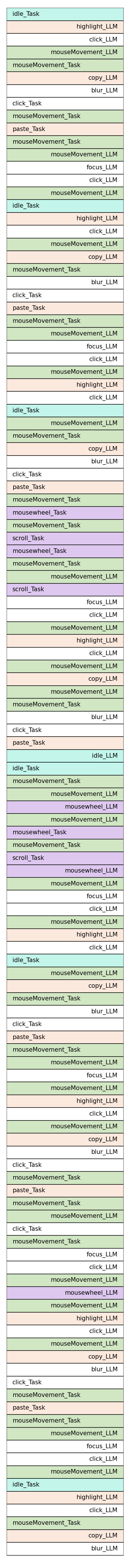}
        \caption{\expThree{}, Window 60, High Overreliance}
        \label{fig:task3_copy_paste_window_60_high_overreliance}
    \end{subfigure}
    \begin{subfigure}[b]{0.2\textwidth}
        \centering
        \includegraphics[width=0.5\textwidth]{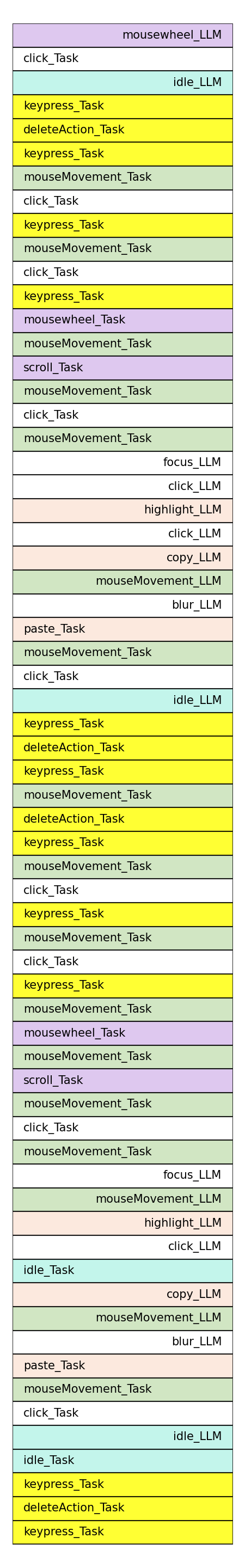}
        \caption{\expThree{}, Window 60, Low Overreliance}
        \label{fig:task3_copy_paste_window_60_low_overreliance}
    \end{subfigure}
    \caption{Copy and Paste for \expThree{}}
    \label{fig:task3_copy_paste_window_60}
\end{figure}

\begin{figure}[htbp]
    \centering
    \begin{subfigure}[b]{0.2\textwidth}
        \centering
        \includegraphics[width=\textwidth]{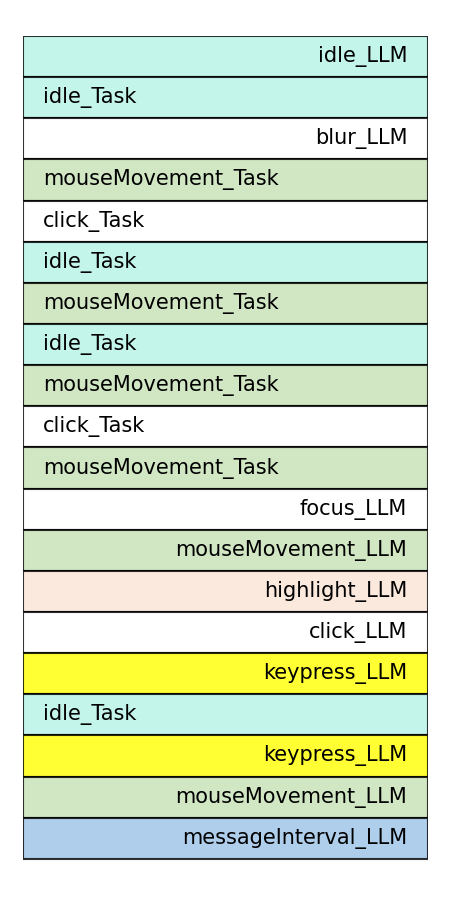}
        \caption{\expOne{}, Window 50, High Overreliance}
        \label{fig:task1_trust_window_30_high_1}
    \end{subfigure}
    \begin{subfigure}[b]{0.2\textwidth}
        \centering
        \includegraphics[width=\textwidth]{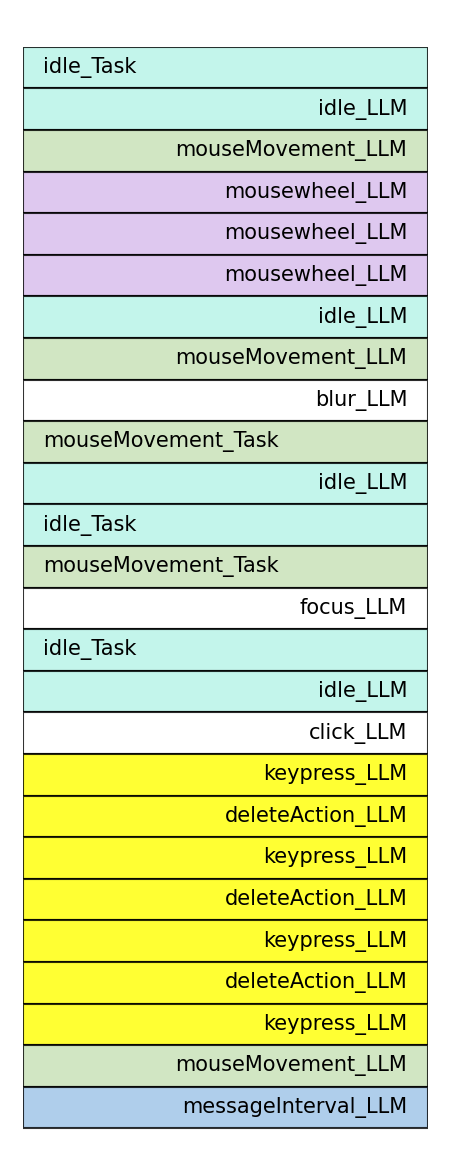}
        \caption{\expOne{}, Window 50, High Overreliance}
        \label{fig:task1_trust_window_30_high_2}
    \end{subfigure}
    \caption{Two Patterns in \expOne{} Trusting \llmres{}}
    \label{fig:task1_trust}
\end{figure}

\begin{figure}[htbp]
    \centering
    \begin{subfigure}[b]{0.2\textwidth}
        \centering
        \includegraphics[width=\textwidth]{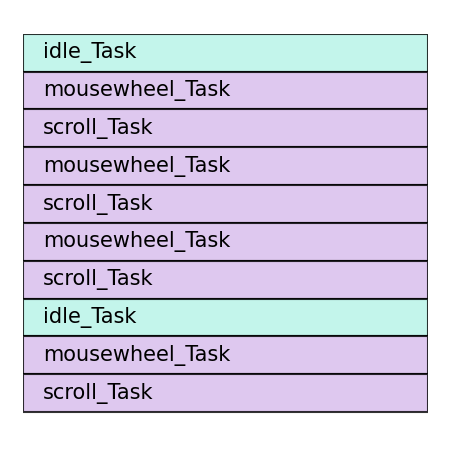}
        \caption{\expTwo{}, Window 10, Low Overreliance}
        \label{fig:task2_consecutive_mousewheel_window_10}
    \end{subfigure}
    \begin{subfigure}[b]{0.2\textwidth}
        \centering
        \includegraphics[width=\textwidth]{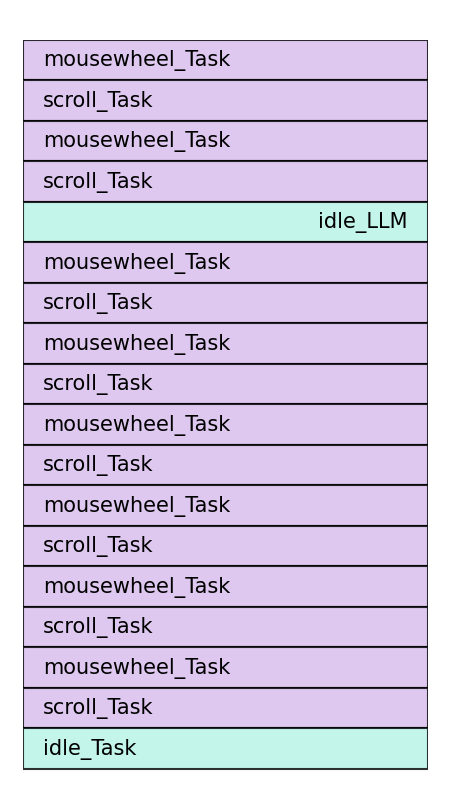}
        \caption{\expTwo{}, Window 20, Low Overreliance}
        \label{fig:task2_consecutive_mousewheel_window_20}
    \end{subfigure}
    \caption{Consecutive Mousewheel for \expTwo{}}
    \label{fig:task2_consecutive_mousewheel}
\end{figure}

Figure~\ref{fig:task3_copy_paste_window_60} corresponds to Section~\ref{subsubsec:copy_paste_frequency}.
Figure~\ref{fig:task2_consecutive_mousewheel} corresponds to Section~\ref{subsubsec:focused_task_comprehension}.
Figure~\ref{fig:task1_interpreting_following} and Figure~\ref{fig:task1_interpreting_following_supplement} corresponds to Section~\ref{subsubsec:pic_interpreting_vs_following}.
Figure~\ref{fig:task2_locating} corresponds to Section~\ref{subsubsec:coarse_vs_fine_grained_locating}.
Figure~\ref{fig:task1_trust} corresponds to Section~\ref{subsubsec:deliberating_vs_wavering}.

\label{Appendix:RawData_from_Behavior_Clustering}

\subsection{Explanation of Each Dimension for Action Vector}
\label{subsec:explanation_vector_dimension}
\begin{enumerate}
    \item[] \textbf{0–14}: One-hot encoding for 15 action types.
    \item[] \textbf{15}: Normalized timestamp.
    \item[] \textbf{16–34}: Action-specific attributes (set to 0 if action event does not match): 
    \begin{enumerate}
        \item[] 21: totalMouseMovement (mouseMovement).
        \item[] 22: deltaY (mousewheel).
        \item[] 23: deltaY (scroll).
        \item[] 24: keyCount (keypress).
        \item[] 25: keyCount (delete).
        \item[] 26: textLength (copy).
        \item[] 27: textLength (paste).
        \item[] 28: textLength (highlight).
        \item[] 29–31: One-hot encoding for mousewheel direction (up, down, stationary).
        \item[] 32–34: One-hot encoding for scroll direction (up, down, stationary).
    \end{enumerate}
    \item[] \textbf{35–36}: One-hot encoding for page type (AI page or tasksheet page).
\end{enumerate}

\subsection{Computing Predictive Score for Test Action Sequence}
\label{appendix:computing_predictive_score_for_test}
Inspired by the regression approach in random forest algorithms \cite{ho1995random}, we designed a method to evaluate the model’s predictive capability for overreliance. 
For each action sequence in the test set, we first computed its corresponding latent vector. Then, we found the top n ($n=5$) nearest neighbor latent vectors in the training set that share the same cluster label. Finally, we averaged the actual overreliance scores of these training set neighbors to obtain the predicted score for the test action sequence.

\subsection{Overreliance Levels}
\label{appendix:overreliance_levels}
Lastly, to better characterize the overreliance scores of each cluster $C_{i}$, we compared the overreliance score of it to the rest of the data $D \setminus C_{i}$ (with $D$ denote the full set of data) within each task-window combination. The results gives three categories:

\begin{enumerate} [leftmargin=0.25in]
    \item High overreliance: $C_{i}$'s score significantly > $D \setminus C_{i}$
    \item Low overreliance: $C_{i}$'s score significantly < $D \setminus C_{i}$
    \item Neutral: no significant difference
\end{enumerate}

\subsection{Questionaires after Each Task}
\label{subsec:questionaire_after_task}
Participants are required to rate from 1 to 7 for the following questions, with 1 indicating ``not agree at all'' and 7 indicating ``extremely agree''.

\begin{enumerate}[leftmargin=*, widest=12, itemsep=0.5em, parsep=0.5em]
    \item The system is deceptive (words related to this: Deception, Lie, Falsity, Betray, Misleading, Phony, Cheat).
    \item The system behaves in an underhanded (sneaky, steal) manner.
    \item I am suspicious of the system's intent, action, or outputs (Mistrust, Suspicion,  Distrust).
    \item I am wary (Beware) of the system.
    \item The system's actions will have a harmful or injurious outcome (Cruel, Harm).
    \item I am confident in the system (Assurance).
    \item The system provides security.
    \item The system has integrity (Honor).
    \item The system is dependable (Fidelity, Loyalty).
    \item The system is reliable (Honesty, Promise, Reliability, Trustworthy, Friendship, Love). 
    \item I can trust the system.
    \item I am familiar with the system.    
\end{enumerate}

Participants are then required to answer the following question:
\begin{enumerate}[leftmargin=*, widest=12, itemsep=0.5em, parsep=0.5em]
    \item In two sentences, explain how you arrived at your final answer. Describe how you utilized the information provided by LLM in reaching your conclusion. 
\end{enumerate}

Finally, participants are required to rate from  1 to 7 for the following questions, with 1 indicating ``not familiar at all'' and 7 indicating ``extremely familiar''.

\begin{enumerate}[leftmargin=*, widest=12, itemsep=0.5em, parsep=0.5em]
    \item How familiar are you with the domain knowledge described in the task?
    \item How motivated are you to complete this task to the best of your ability?
\end{enumerate}

\subsection{Analysis Pipeline Details}
\label{appendix:analysis_pipeline}

This appendix provides additional technical details referenced in Section~\ref{sec:analysis_method}, including preprocessing procedures, feature encoding, model architecture, clustering configuration, and validation mechanisms used in the analysis pipeline.

\subsubsection{Preprocessing Methodology}
\label{appendix:preprocessing}

The raw interaction logs were first cleaned to exclude participants whose data were incomplete or corrupted due to upload issues or browsing errors. This yielded 62, 70, and 60 valid logs in \expOne{}, \expTwo{}, and \expThree{}, respectively. The remaining logs were merged across the Task Page and LLM Page by aligning all events chronologically using their start timestamps and labeling their page origin.

To eliminate noise and improve interpretability, we aggregated low-level events into higher-level action units based on temporal proximity and semantic similarity. For instance, multiple cursor movements within 200 milliseconds or continuous key presses were merged into a single action event. Each action was enriched with metadata including its duration, position on the page, and contextual type. Time was normalized such that the first observable action had a timestamp of 0.0, and the full 90-minute session duration was scaled to the range [0, 1.0]. Each action was assigned a unique sequential ID representing its temporal ordering within the session.

\subsubsection{Action Vector Encoding}
\label{appendix:vector_dimensions}

Each action sequence was transformed into a sequence of 37-dimensional vectors. These vectors included four main components: one-hot encoded action type (15 dimensions), a normalized timestamp (1 dimension), a one-hot encoded page context (Task Page or LLM Page, 2 dimensions), and a set of 19 additional features specific to the action type. These included continuous values such as cursor movement distance, scroll amount, prompt length, and burst typing duration, as well as categorical variables such as input modality or mouse wheel direction. Continuous attributes were log-transformed prior to training to mitigate skew caused by outliers. Categorical variables were encoded using one-hot or binary representations, depending on cardinality.

A complete list of encoded action attributes is summarized in Table~\ref{tab:action_feature_encoding}. Vectorized sequences were used as direct input to the embedding model described in the next section.

\begin{table}[h]
\centering
\caption{Summary of Encoded Action Features}
\label{tab:action_feature_encoding}
\scriptsize
\begin{tabular}{p{0.25\linewidth} p{0.45\linewidth} p{0.2\linewidth}}
\toprule
Feature & Description & Number of Dimension \\
\midrule
\texttt{action\_type} & One of 15 categories (e.g., scroll, click, type, prompt-submit) & 15\\
\texttt{timestamp} & Normalized session time (0.0 to 1.0) & 1\\
\texttt{page\_context} & Task page or LLM page & 2\\
\texttt{total\_mouse\_movement} & Total distance moved by \textit{mouseMovement} in physical pixels & 1\\
\texttt{mouse\_movement\_duration} & The duration of a continuous \textit{mouseMovement} & 1\\
\texttt{scroll\_duration} & The duration of a continuous \textit{scroll} or \textit{mouseWheel} & 1\\
\texttt{scroll\_distance} & Total distance moved by \textit{scroll} in physical pixels & 1\\
\texttt{mouseWheel\_distance} & Total distance moved by \textit{mouseWheel} in physical pixels & 1\\
\texttt{scroll\_direction} & \textit{scroll} up, down or none & 3\\
\texttt{mouseWheel\_direction} & \textit{mousewheel} up, down or none & 3\\
\texttt{keypress\_duration} & The duration of a continuous \textit{keypress} input & 1\\
\texttt{keypress\_keyCount} & The total count of a continuous \textit{keypress} input & 1\\
\texttt{copy\_textLength} & The total length of copied text & 1\\
\texttt{paste\_textLength} & The total length of pasted text & 1\\
\texttt{highlight\_textLength} & The total length of highlighted text & 1\\
\texttt{delete\_duration} & The duration of a continuous \textit{delete} action & 1\\
\texttt{delete\_keyCount} & The total count of a continuous \textit{delete} input & 1\\
\texttt{idle\_duration} & The duration of a continuous \textit{idle} & 1\\
\bottomrule
\end{tabular}
\end{table}

\subsubsection{Autoencoder Configuration}
\label{appendix:autoencoder_architecture}

To generate a fixed-dimension representation of variable-length interaction sequences, we used a Transformer-based autoencoder model. Each preprocessed action sequence was prepended with a special \texttt{[CLS]} token initialized to a constant value. The encoder consisted of three Transformer layers, each with four self-attention heads, and intermediate feedforward layers of size 128. The encoder output corresponding to the \texttt{[CLS]} token (size 64) was used as the final embedding of the entire sequence.

The decoder network mirrored the encoder structure, reconstructing the original sequence from the compressed embedding. Training was optimized with a combination of three loss terms: (1) categorical cross-entropy for recovering the original action type, (2) binary cross-entropy for page classification, and (3) a weighted reconstruction loss combining mean squared error (for continuous features) and cross-entropy (for categorical ones). Training used the Adam optimizer with a learning rate of 1e-4 and a batch size of 32. We trained a separate model for each task and window size, resulting in 18 autoencoders (3 tasks × 6 window sizes). Validation was conducted using a leave-one-participant-out setup to ensure generalizability, and models were early-stopped upon convergence.

\subsubsection{Clustering Procedure}
\label{appendix:clustering_details}

After embedding, each segmented sequence was represented by a 64-dimensional vector corresponding to the \texttt{[CLS]} output of the autoencoder. These latent vectors were clustered using the DBSCAN algorithm to discover commonly recurring patterns of behavior. To improve robustness and avoid parameter sensitivity, we performed multiple DBSCAN runs using a grid search over parameters: the neighborhood distance threshold (\texttt{eps}) was varied from 0.2 to 1.0 (step size 0.1), and the minimum required cluster size (\texttt{min\_samples}) was varied from 3 to 10. Clusters that appeared in at least three different parameterizations were considered stable and retained for downstream analysis.

To assign clusters to unseen test points, we used a $k$-nearest neighbor classifier ($k=5$) trained on the latent vectors and their corresponding cluster assignments from the training set. This allowed consistent cluster membership inference while preserving training/test separation.

\subsubsection{Validation of Cluster Utility}
\label{appendix:computing_predictive_score_for_test}

After clustering, we validated each cluster using two criteria designed to ensure that the behavior it represented was both internally consistent and predictive of user overreliance.

First, for internal consistency (intrinsic similarity), we compared the distribution of overreliance scores for sequences in the training and test sets that had been assigned to the same cluster. We performed independent two-sample t-tests for each cluster. Clusters where the training and test scores were not significantly different ($p > .05$) were retained as stable across users.

Second, for predictive capability, we compared the mean overreliance score of the training sequences assigned to a cluster with the mean score of test sequences assigned to that cluster. We retained clusters in which the absolute difference between these two means was below a fixed threshold $\delta = 0.15$, empirically tuned to balance sensitivity with robustness.

\subsubsection{Overreliance Score Stratification}
\label{appendix:overreliance_levels}

To simplify interpretation, we discretized the continuous overreliance scores into three ordinal levels: high, neutral, and low. Users scoring in the top quartile within each task’s overreliance distribution were labeled as high-overreliance, while those scoring in the bottom quartile were labeled as low-overreliance. Those with mid-range scores were labeled neutral. These labels were not used for inference or clustering but aided in interpreting cluster characteristics in qualitative analysis.

\smallskip

The methods described here provide the technical foundation for the findings reported in Section~\ref{sec:quantitative_findings}, where we qualitatively interpret representative clusters to shed light on behavioral patterns that align with LLM overreliance dynamics.

\subsection{Pilot Study}
\label{subsec:pilot_study}

Over time, we conducted multiple rounds of pilot studies with a total of 21 participants (9F, 12M). These pilots allowed us to testing different methods for capturing cognitive ground truth, calibrate task difficulty and constraints, and evaluate different categories of tasks.

\end{document}